\documentclass[11pt,a4paper]{article}
\usepackage{jcappub}

\usepackage{epstopdf}
\usepackage{natbib}

\newcommand{\etal}{{et\ al.}}

\def\aa{{\sl Astron.\ \&\ Astrophys.\ }}

\def\apj{{\sl Astrophys.\ J.\ }}
\def\apjl{{\sl Astrophys.\ J.\ Lett.\ }}
\def\apjs{{\sl Astrophys.\ J.\ Supp.\ }}

\def\jcap{{\sl J.\ Cosm.\ Astroparticle\ Phys.\ }}

\def\mnras{{\sl MNRAS\ }}

\def\physrep{Phys. Rept.\ }

\def\prd{{\sl Phys.\ Rev.\ D\ }}
\def\prl{{\sl Phys.\ Rev.\ Lett.\ }}

\title{Measuring primordial gravitational waves from CMB $B-$modes in cosmologies with generalized expansion histories}

\author[a,b]{Claudia Antolini,}
\author[a,b]{Matteo Martinelli,}
\author[a,b]{Yabebal Fantaye,}
\author[a,b]{Carlo Baccigalupi}
\affiliation[a]{SISSA, Via Bonomea 265, Trieste, I-34136, Italy}
\affiliation[b]{INFN, Sezione di Trieste, Via Valerio 2, 34127 Trieste, Italy}
\emailAdd{claudia.antolini@sissa.it}
\emailAdd{mmartin@sissa.it}
\emailAdd{fantaye@sissa.it}
\emailAdd{bacci@sissa.it}
\abstract{ We evaluate our capability to constrain the abundance of primordial tensor perturbations (primordial gravitational waves, PGWs) in cosmologies with generalized expansion histories in the epoch of cosmic acceleration. Forthcoming satellite and sub-orbital experiments probing polarization in the Cosmic Microwave Background (CMB) are expected to measure the $B-$mode power in CMB polarization, coming from PGWs on the degree scale, as well as gravitational lensing on arcminute scales; the latter is the main competitor for  the  measurement of PGWs, and is directly affected by the underlying expansion history, determined by the presence of a Dark Energy (DE) component. In particular, we consider early DE possible scenarios, in which the expansion history is substantially modified at the epoch in which the CMB lensing is most relevant.  We show that {the introduction of a parametrized DE } may induce a variation as large as  $30\%$ in the ratio of the power of lensing and PGWs on the degree scale. We 
find that adopting the nominal specifications of upcoming satellite measurements, the constraining power on PGWs is weakened by the inclusion of the extra degrees of freedom, resulting in a reduction of about $10\%$ of the upper limits on $r$ in fiducial models with no GWs, as well as a comparable increase in the error bars in models with non-zero tensor power. Moreover, we find that the inclusion of sub-orbital CMB experiments, capable of mapping the $B-$mode power up to the angular scales which are affected by lensing, has the effect of restoring the forecasted performances with a fixed cosmological expansion history corresponding to a cosmological constant. Finally, we show how the combination of CMB data with Type Ia SuperNovae (SNe), Baryonic Acoustic Oscillations (BAO) and Hubble constant allows to constrain simultaneously the primordial tensor power and the DE quantities in the parametrization we consider, consisting of present abundance and first redshift derivative of the energy density. { We 
compare this study with results obtained using the forecasted lensing potential measurement precision from CMB satellite observations, finding consistent results.}}
\arxivnumber{1208.3960}

\begin{document}

\maketitle

\section{Introduction}
\label{sec:introduction}

Anisotropies in the Cosmic Microwave Background (CMB) represent one of the pillars of modern cosmology. Their statistical distribution, characterized primarily 
by the angular power spectrum, is consistent with a flat Friedmann Robertson Walker metric, expanding at a Hubble rate corresponding to about 70 km/s/Mpc, 
and composed by three main cosmological components, namely baryons and leptons representing about $4\%$ of the total energy density, dark matter (DM, about $21\%$) 
constituting the large part of the gravitational potential around collapsed or forming cosmological structures, and about $75\%$ of a Dark Energy (DE) component, 
similar or coincident with a Cosmological Constant (CC), responsible for a late time phase of accelerated expansion. The primordial spectrum of density perturbations 
is almost scale invariant, corresponding to a Harrison-Zel'dovich power law shape in wavenumbers. 
Three satellites have been observing CMB anisotropies, the Cosmic Background Explorer \citep{smoot_etal_1992}, the Wilkinson Microwave Anisotropy Probe \citep{komatsu_etal_2011}, 
and Planck, which is expected to release cosmological data in early 2013 \citep{the_planck_collaboration_2011}. Space observations will provide an all sky measurement of total intensity 
and polarization anisotropies down to a resolution of a few arcminutes, and a sensitivity of a few $\mu$K per resolution element. 

A number of sub-orbital experiments are planned and have been observing selected regions of the sky and frequency spectrum, looking for arcminute and sub-arcminute scale anisotropies 
in total intensity ($T$), as well as polarization\footnote{see NASA ADS for the list of operating or planned sub-orbital CMB experiments.}. These observations will target most important and 
yet still undetected effects, dominating the curl component ($B-$modes) of the linear polarization pattern in CMB anisotropies \citep{kamionkowski_etal_1997,zaldarriaga_seljak_1997}. 
On arcminute angular scales, the latter are dominated by the gravitational lensing of the anisotropies at last scattering by means of forming cosmological structures along the line of sight. 
A fraction of the gradient component of polarization ($E-$modes), dominating because powered by density fluctuations responsible for sub-degree acoustic oscillations at last scattering, 
is converted into $B-$modes by means of gravitational lensing \cite{zaldarriaga_seljak_1998}. 
The power spectrum of the underlying DM distribution, and the primordial $E-$modes, produce a characteristic and broad lensing peak centered at $l\simeq 1000$ in the $B-$mode power spectrum. 
Gravitational lensing has been recently detected in the damping tail of $T$ anisotropies by several groups \cite{keisler_etal_2011,hlozek_etal_2012}, also cross-correlating the lensing with 
observed structures \citep{sherwin_das_2012}, while  $B-$modes have not yet been detected, see \citet{the_quiet_collaboration_2011} for the current upper limits. 
On the degree angular scales on the other hand, a primordial spectrum of tensor anisotropies or cosmological Gravitational Waves (GWs) would produce a narrow peak, rapidly vanishing on 
sub-degree angular scales, not supported by radiation pressure from massive particles, as is instead the case for $T$ and $E-$modes. On large angular scales, corresponding to several 
degrees in the sky, the decay of the GWs tail in the $B-$modes can be re-amplified though re-scattering onto electrons in the epoch of cosmic reionization. As for the case of lensing, only upper limits 
exist for the amplitude of PGWs through direct measurement of $B-$modes. The two effects compete for detection, and their different origin, primordial and linear for GWs, 
late and second order for lensing, has been exploited for designing separation techniques \cite{hirata_seljak_2004}. {Furthermore, it has been analysed in the past how an accelerated expansion modifies the shape of the spectrum of PGWs as a result of propagation in a different space-time \cite{zhangetal}. }

The lensing peak of $B-$mode anisotropies strongly depends on the history of cosmic expansion. It has been shown \cite{acquaviva_baccigalupi_2006} that its amplitude may undergo variations of 
order $10\%$ if the DE is dynamical at the epoch corresponding to the onset of acceleration, i.e. about $z\simeq 1$, in which its actual amplitude is poorly constrained by existing 
measurements of the CMB or large scale structures. The $B-$mode lensing peak as a DE probe has been investigated by several authors \citep{acquaviva_baccigalupi_2006,hu_etal_2006}, who in 
particular have shown how the lensing is capable of breaking the projection degeneracy affecting CMB anisotropies at the linear level, as it was recently confirmed in the context of lensing 
detection for sub-orbital $T-$mode experiments \citep{sherwin_etal_2011}. On the other hand, the detection thresholds for cosmological GWs as well as the accuracy on DE constraints 
from CMB observations have  never been given by taking into account the full set of degrees of freedom, represented not only by the amplitude of primordial GWs, but also by those related to the expansion 
history,  parametrized through suitable DE models. The release of the latter degrees of freedom in the context of experiments aiming at the detection and characterization of $B-$mode anisotropies 
is expected to have a direct impact in the quoted detection thresholds of primordial GWs. 

In this work we explore this issue, by investigating the sensitivity of forthcoming $B-$mode probes on primordial GWs abundance as well as DE dynamics when all the physical degrees of freedom shaping 
the $B-$mode power spectrum are considered and treated jointly. In this context, we consider in particular the interplay { between satellite measurements, accessing large scale polarization and extracting lensing mainly from $T$ and $E$ measurements, and the case of sub-orbital ones, directly probing lensing $B-$modes.} We will take as reference two among the most important forthcoming $B-$mode probes, EBEX \citep{reichborn_kjennerud_etal_2011} and PolarBear \cite{polarbear} as well as the all sky measurements featuring the nominal capabilities from Planck \cite{:2006uk}. 

This work is organized as follows. In Section \ref{sec:spectrum} we describe the impact of a modified expansion history on the CMB lensing power. In Section \ref{sec:analysis} we describe our 
set of simulated data as well as the reference experiments we consider. In Section \ref{sec:results} we show and discuss our results, while in Section \ref{sec:discussion} we draw our 
conclusions. 

\section{Generalized expansion histories: how lensing affects the CMB spectra}
\label{sec:spectrum}

\begin{figure}[t]
\begin{center}
\includegraphics[width=7.5cm]{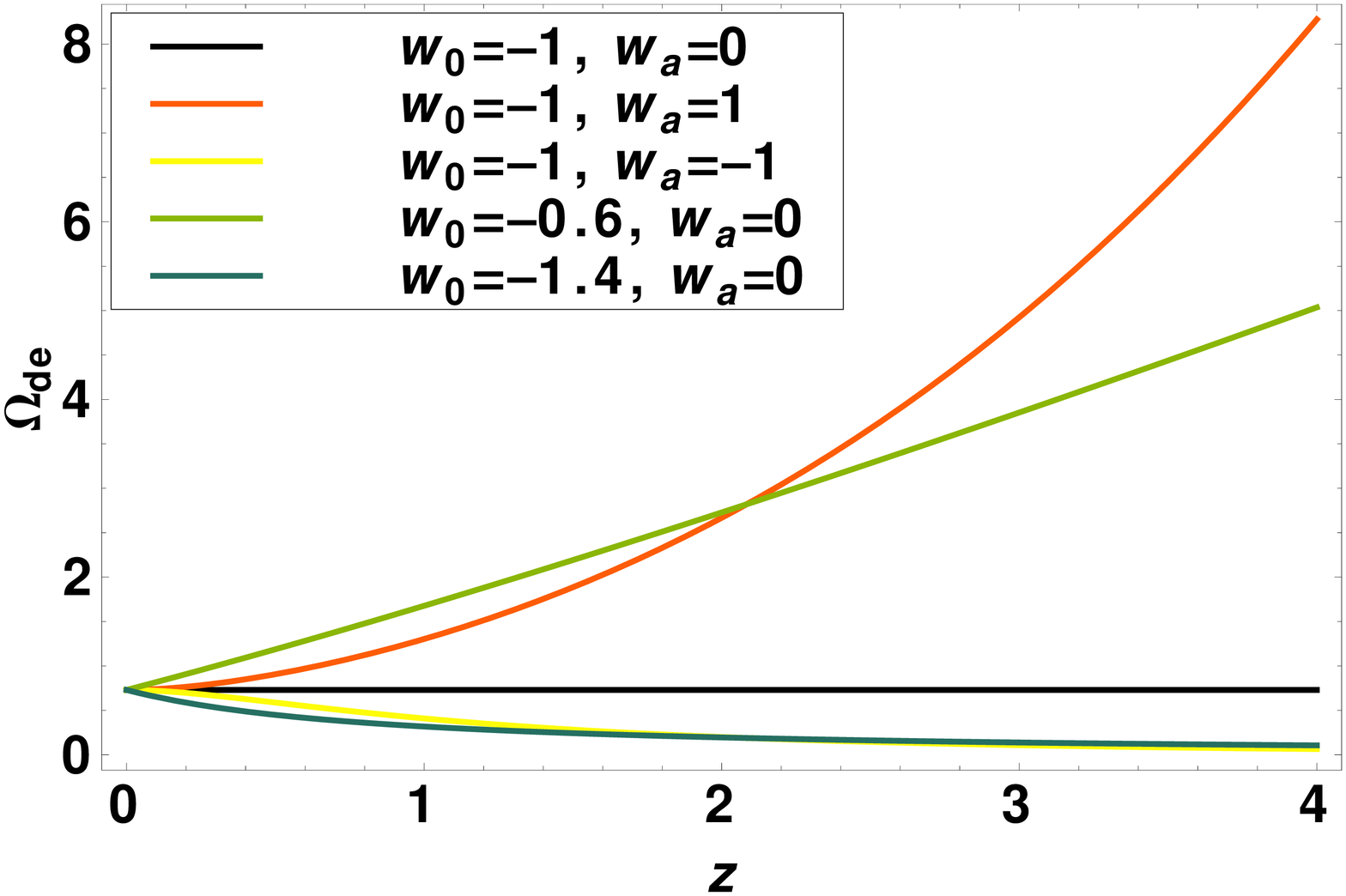}
\includegraphics[width=7.5cm]{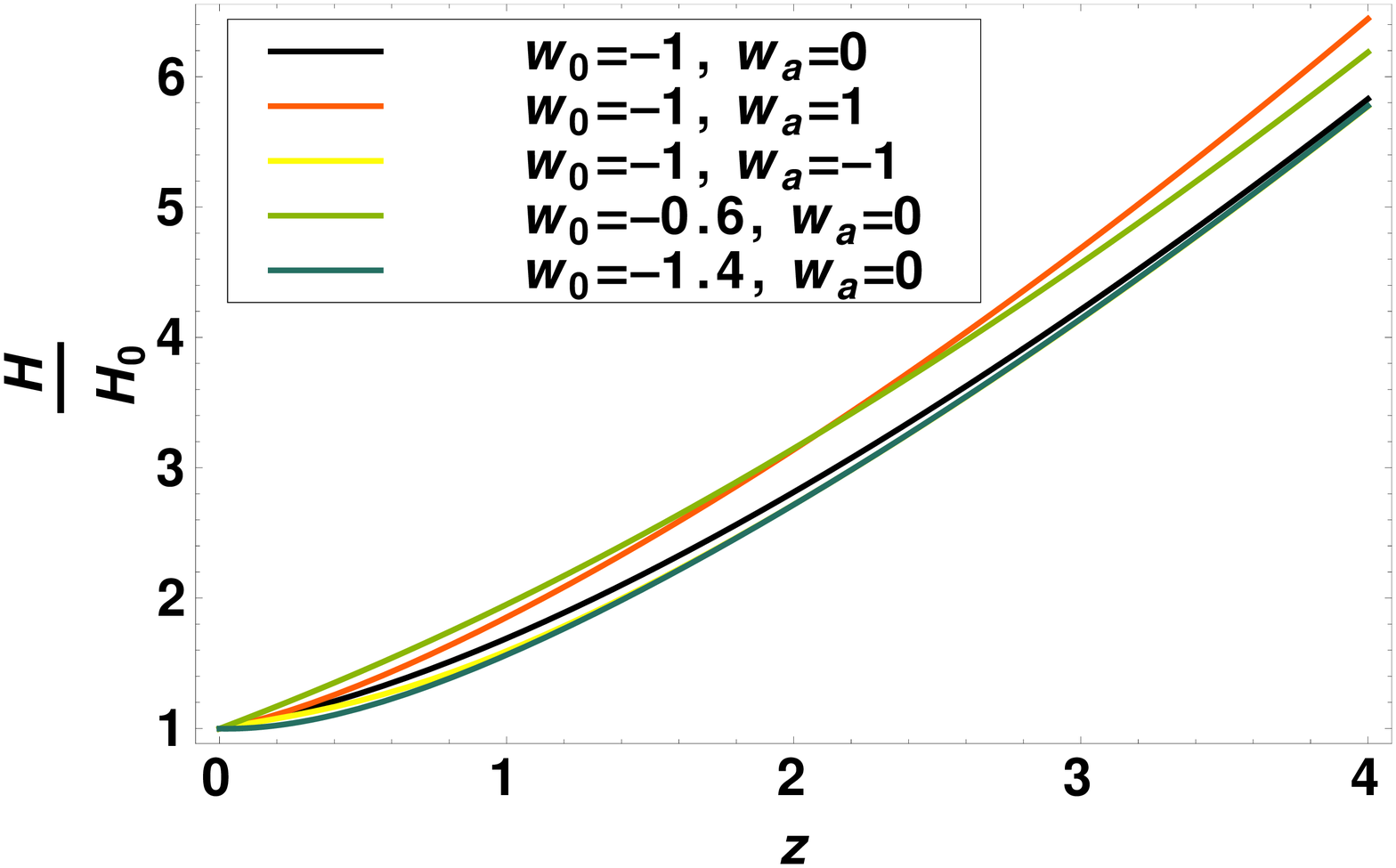}
\vspace{-1cm}
\caption{Left panel: redshift evolution of the DE component with different values of $w_0, w_a$. As the sum $w_0, w_a$ get above $-1$, the DE term becomes increasingly important in the past. 
Right panel: corresponding evolution of the Hubble parameter with redshift with the same expansion histories considered in the left panel.}
\label{fig:DEtime}
\end{center}
\end{figure}

 In this work we consider models of expansion history corresponding to a Cosmological Constant (CC) and its generalization through the equation of state $w=p/\rho$ of the DE evaluated at 
present, as well as its first derivative in the scale factor \citep{CP,Linder}, often labelled CPL. In this modelisation, the DE equation of state and the 
ratio $\Omega_{DE}$ of its energy density with respect to the cosmological critical density are given by 
\begin{equation}\label{eq:evolution}
p= \left[w_0 + (1-a)w_a \right] \rho\ \ ,\ \ 
\Omega_{DE}(z)=\Omega_{DE,0}e^{3\int^z_0 dz^{\prime}
\frac{1 + w(z^{\prime})}{1 + z^{\prime}}}\ .
\end{equation}
Such a parametrization allows for a large set of dynamics in the cosmic acceleration, and in particular an increased DE abundance at the equivalence with cold DM and the onset of acceleration. 
In the following we will see how the evolution of DE with time affects the CMB lensing because of its influence on the structures generating the gravitational potential  responsible 
for the deflection. In Fig. \ref{fig:DEtime}, left panel, one can see how the DE density evolves with time as the $w_0,w_a$ parameters vary. 
In order to get a glimpse on how the lensing process is modified by different expansion histories, let's look again at Eq. (\ref{eq:evolution}) and consider how this influences the evolution with 
redshift of the Hubble parameter $H(z)$, which we can see in Fig. \ref{fig:DEtime} (right panel). 

\begin{figure}[t]
\begin{center}
\includegraphics[width=7.5cm,height=4.8cm]{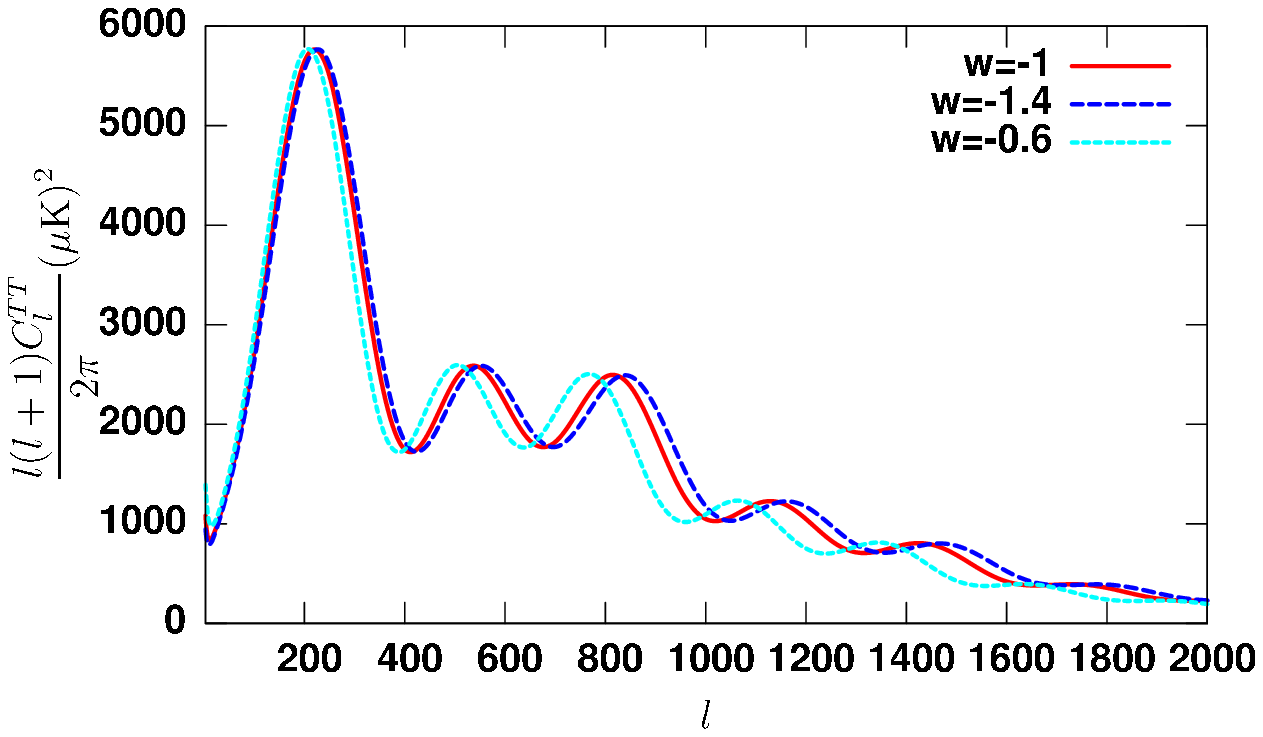}
\includegraphics[width=7.5cm,height=4.8cm]{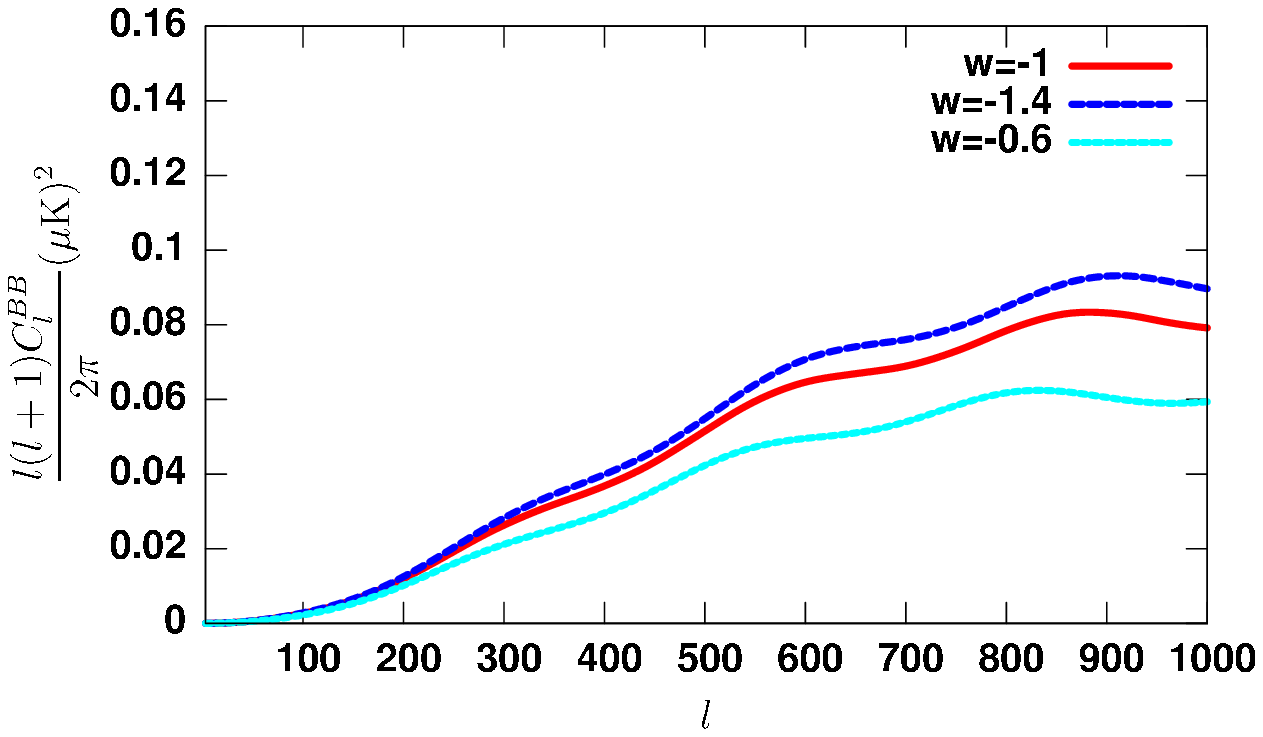}
\caption{Left panel: Variation of the $T-$mode spectrum with different values of $w$.  Right panel: Variation of the $B-$mode spectrum with different values of $w$. }
\label{fig:deg}
\end{center}
\end{figure}

\begin{figure}[t]
\begin{center}
\includegraphics[width=10cm]{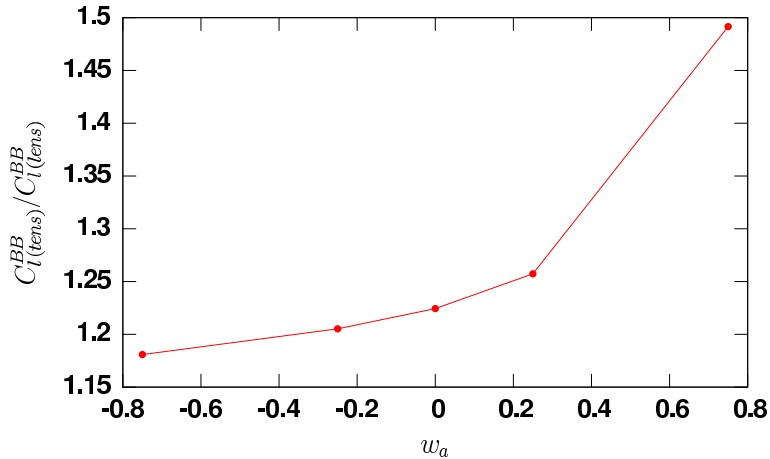}
\caption{Ratio between the primordial $B-$modes ($r=0.05$) and lensing generated $B-$modes at $\ell=100$ with different expansion histories with $w_0$ fixed to -1.}
\label{fig:ratio}
\end{center}
\end{figure}

{Gravitational lensing deflection angle is related to the lensing projected potential $\phi$ (see e.g. \cite{Perotto:2006rj,Calabrese:2009tt}) through the relation
\begin{equation}
d_l^m=-i\sqrt{l(l+1)}\phi_l^m.
\label{eq:defd}
\end{equation}
It is characterized by the lensing deflection power spectrum $C_\ell^{dd}$, which is defined through the ensemble average 
\begin{equation}
 \langle d(a,b)_L^{M*} d(a',b')_{L'}^{M'}\rangle\equiv \delta_{L}^{L'}\delta^{M'}_{M}(C_L^{dd}+N_L^{aa'bb'})~.
\end{equation}
Following \cite{lensextr}, the lensing deflection angle can be inferred by the observed CMB anisotropies through 
\begin{equation}
d(a,b)_L^M=n_L^{ab}\sum_{ll'mm'}W(a,b)_{l\ l'\ L}^{mm'M}a^m_lb^{m'}_{l'}~,
\label{eq:estimator}
\end{equation}
where $a,b$ are the CMB modes $T,\ E,\ B$ modes, $n_L^{ab}$ is a normalization factor introduced to obtain an unbiased estimator and $W(a,b)$ is a weighting factor which leads to the noise $N_L^{aa'bb'}$ on the power spectrum \footnote{We will specify the extraction method followed here (and therefore our choice of $W$) in the next section.}.\\}
We now describe from a physical point of view the CMB lensing process and its sensitivity to the underlying expansion history. For a full mathematical treatment we refer to earlier works \citep{hu2000,bartelmann_schneider_2001,weaklens}. As the Hubble expansion rate grows in the past  with respect to $\Lambda$CDM, the cosmic expansion rate increases.
Its value at the epoch of structure formation will determine how efficient the process of  structure formation is, and consequently the abundance of available lenses: 
the lower is the Hubble rate in that epoch, the lower the friction represented by the expansion with respect to structure formation, the higher the number of lenses will be. As noticed 
by \citet{acquaviva_baccigalupi_2006}, the latter occurrence is rather sensitive to the DE abundance at the epoch at which lensing is most effective, $z\simeq 1\pm 0.5$, 
and rather independent of the DE properties at earlier and later epochs than that, simply because by geometry, the lensing cross section peaks about halfway between sources and 
observer. 
\begin{figure}[ht]
\begin{center}
\includegraphics[width=10cm]{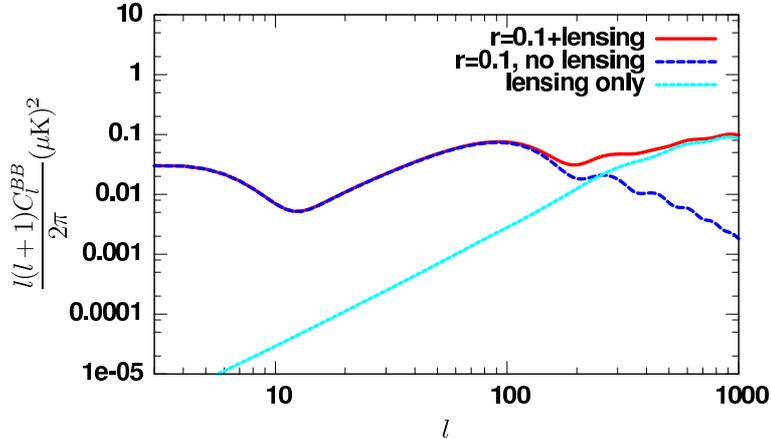}
\caption{$B-$modes for CMB polarization anisotropies with different contributions given by primordial tensor modes only with $r=0.1$ (green), by lensing only (blue), and the total 
for both lensing and $r=0.1$ tensor modes.}
\label{fig:unlensed}
\end{center}
\end{figure}
 The distribution of lenses, following the power spectrum of density perturbations, as well as the geometrical properties mentioned above, determine the efficiency of CMB lensing to 
peak on arcminute angular scales, corresponding to structures from a few to about $10^{2}$ comoving Mpc. Being a non-linear effect, lensing redistributes primordial anisotropy power 
of single multipoles at last scattering on a finite interval of scales. The net effect on $T$ and $E$ is a smearing of acoustic peaks and the dominance in the damping tail region, 
corresponding to multipoles of $\ell \gtrsim 1000$, where primordial anisotropies die out because of diffusion damping, and the only power comes from larger scales because of lensing. 
As we already discussed, for $B-$modes the effect is rather different. In Fig. \ref{fig:unlensed} we show the various contributions to $B-$modes, coming from primordial GWs on 
degree and super-degree angular scales, and from lensing on arcminute ones. The latter effect arises because a fraction of $E-$modes is transferred to $B$ because of the deflection itself. The sensitivity 
of this process to the underlying DE properties is described in Fig. \ref{fig:deg}, where the $T$ and $B$ spectra are shown for various cases. The geometric shift in $T$ is due to the change in 
comoving distance to the last scattering, given by 
\begin{equation}
\label{dls}
d_{LS}= H^{-1}_0 \int^{z_{LS}}_0 dz\left[\Omega_m (1+z)^3+\Omega_{DE,0}e^{3\int^z_0 dz^{\prime}\frac{1 + w(z^{\prime})}{1 + z^{\prime}}}\right]^{-1/2} 
\end{equation}
where $H_{0}$ is the Hubble parameter, $\Omega_{m}$ is the matter abundance today relative to the critical density and the contributions from radiation and curvature are neglected. Clearly, 
the same value of $d_{LS}$ can be obtained with various combinations of parameters, including the DE, creating the so called projection degeneracy, already addressed in \cite{acquaviva_baccigalupi_2006}. The lensing, for $B-$modes in particular, 
shown in the right panel, is capable of breaking it, because of its sensitivity to the DE abundance at the epoch in which its cross section is non-zero. Indeed, 
looking again at Fig. \ref{fig:DEtime}, we  see that the DE density at the epoch we are considering follows an opposite behaviour with 
respect to the curves represented in Fig. \ref{fig:deg}: the lower the curve, the higher the value of the expansion rate at the relevant epoch for lensing leading to an increasing suppression 
of the power, the higher the dark energy density, as already discussed above. 
It is already well known \cite{hirata_seljak_2004a} that the gravitational lensing signal constitutes a  fundamental contaminant in the PGWs spectrum. The latter is parametrized 
by the ratio between the tensor and scalar power in the primordial perturbation power spectra, $r$. As for scalars, the power spectrum of PGWs is also characterized by a spectral index. We work here in the hypothesis of single field inflationary models, which relate the tensor spectral index to $r$, without introducing any additional parameter; a discussion on parameter estimation without this assumption may be found in  \cite{Efstathiou:2001cv, melchiorri}.\\
Our aim in this  work is trying to infer how a simultaneous constraint can be affected 
by the presence of both signals in data, and in particular to determine the degradation, if any, of the constraint on $r$ as the background expansion is allowed to vary according to a CPL 
parametrization. As we have seen, this heavily affects the lensing peak of the CMB:  for a better quantification of this point, we show in Fig. \ref{fig:ratio} how the ratio of the two contributions 
at the peak of the GWs power, corresponding to $\ell\simeq 100$, can vary macroscopically because of the variation in the DE dynamics, reaching $50\%$. It is clear that it is necessary to study 
the parameter space represented by $r,\, w_0, \,w_a$ jointly, in order to understand the constraining power based on data on CMB $B-$modes.

\section{Simulated data and analysis}
\label{sec:analysis}

In this Section we describe our methodology related to the simulation of CMB data as well as its analysis.
{In order to obtain a forecast for different parameters using nominal instrumental performances, a Fisher matrix approach is often adopted for estimating covariances. However, the latter approach is rigorously valid only if the likelihood shape of parameters is Gaussian. In our case, as we will show, the shape of the likelihood for $r$ deviates substantially from a Gaussian; in order to avoid inaccuracies, as it was pointed out in recent works \cite{wolz} we prefer to avoid such a simplification. Another reason for doing so is that we will make use of different datasets in our analysis, described later in this Section, and we cannot assume that no degeneracies will arise from this combination. For these reasons, our approach consists in computing the full likelihood shapes by using a Markov chains approach. } We exploited extensively the publicly available 
software package {\tt cosmomc}\footnote{\tt cosmologist.info} for Markov Chain Monte Carlo (MCMC) analysis of CMB datasets \citep{Lewis:2002ah}. \\
We create simulated CMB datasets for $T$, $E$ and $B-$modes, adopting the specifications of Planck \cite{the_planck_collaboration_2011}, EBEX \citep{reichborn_kjennerud_etal_2011} and PolarBear \citep{polar} 
experiments. In Table \ref{tab:exp} we list the relevant parameters adopted in the present work. The fiducial model for the standard cosmological parameters is the best fit from the WMAP 
seven years analysis \citep{komatsu_etal_2011}, concerning flat $\Lambda$CDM parametrizing {the abundances of CDM and baryons plus leptons ($h^{2}\Omega_{c}$, 
$h^{2}\Omega_{b}$, respectively), $100 \cdot \theta$, where $\theta$ is the ratio of the sound horizon to the angular diameter distance, the optical depth $\tau$ of cosmological 
reionization, the spectral index $n_{s}$ and amplitude $A_{s}$ of the primordial power spectrum of density perturbations, the parameters for evolving DE $w_0, \, w_a$}. In the present work we want to study the effects that a generalized expansion history has on the cases of a null as well as a positive detection of $r$. { In Table \ref{tab:wmap} the values used to compute the simulated spectra are shown.}

\begin{figure}
\begin{center}
\includegraphics[width=7cm,height=5cm]{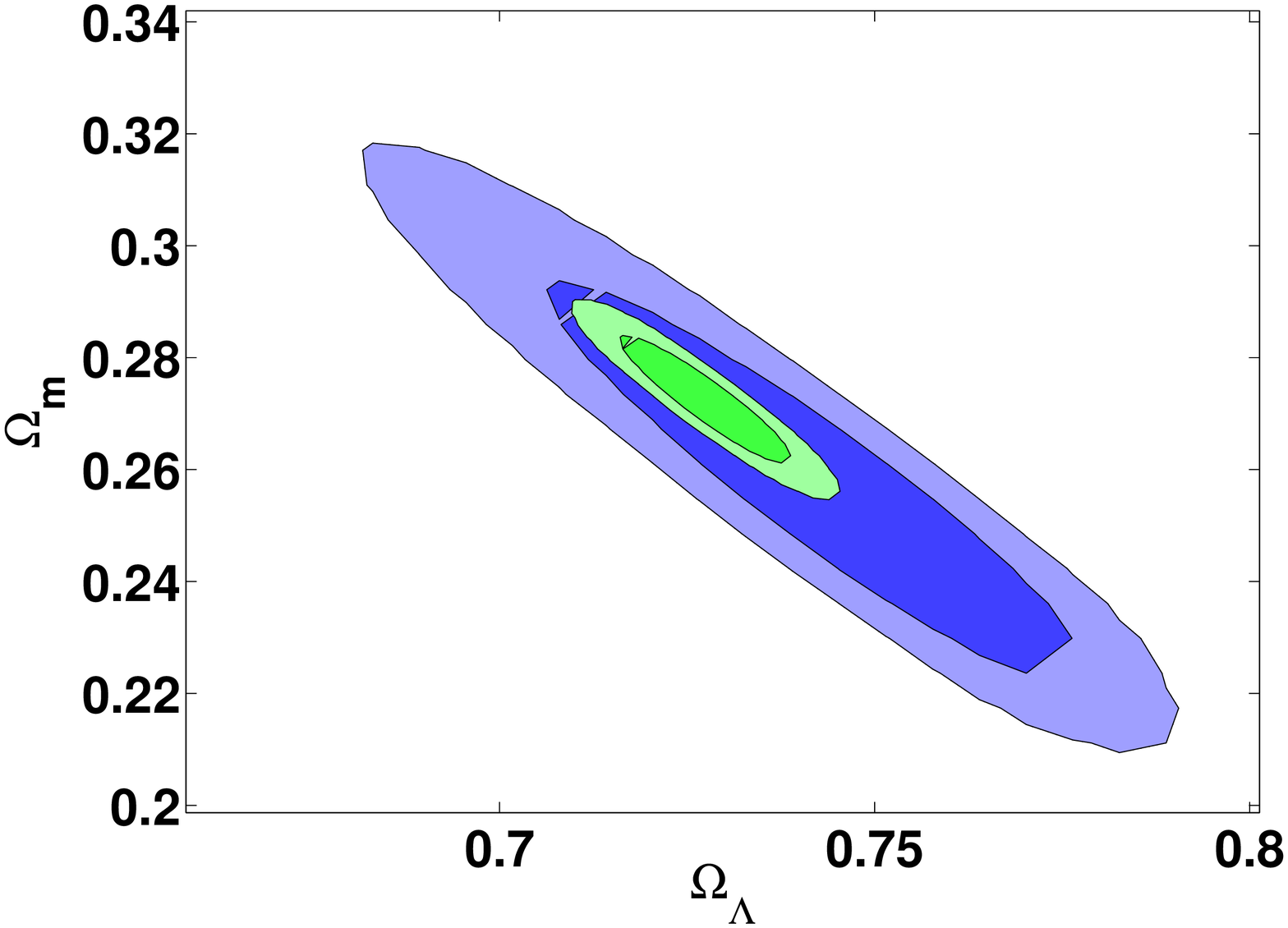}
\includegraphics[width=7cm,height=5cm]{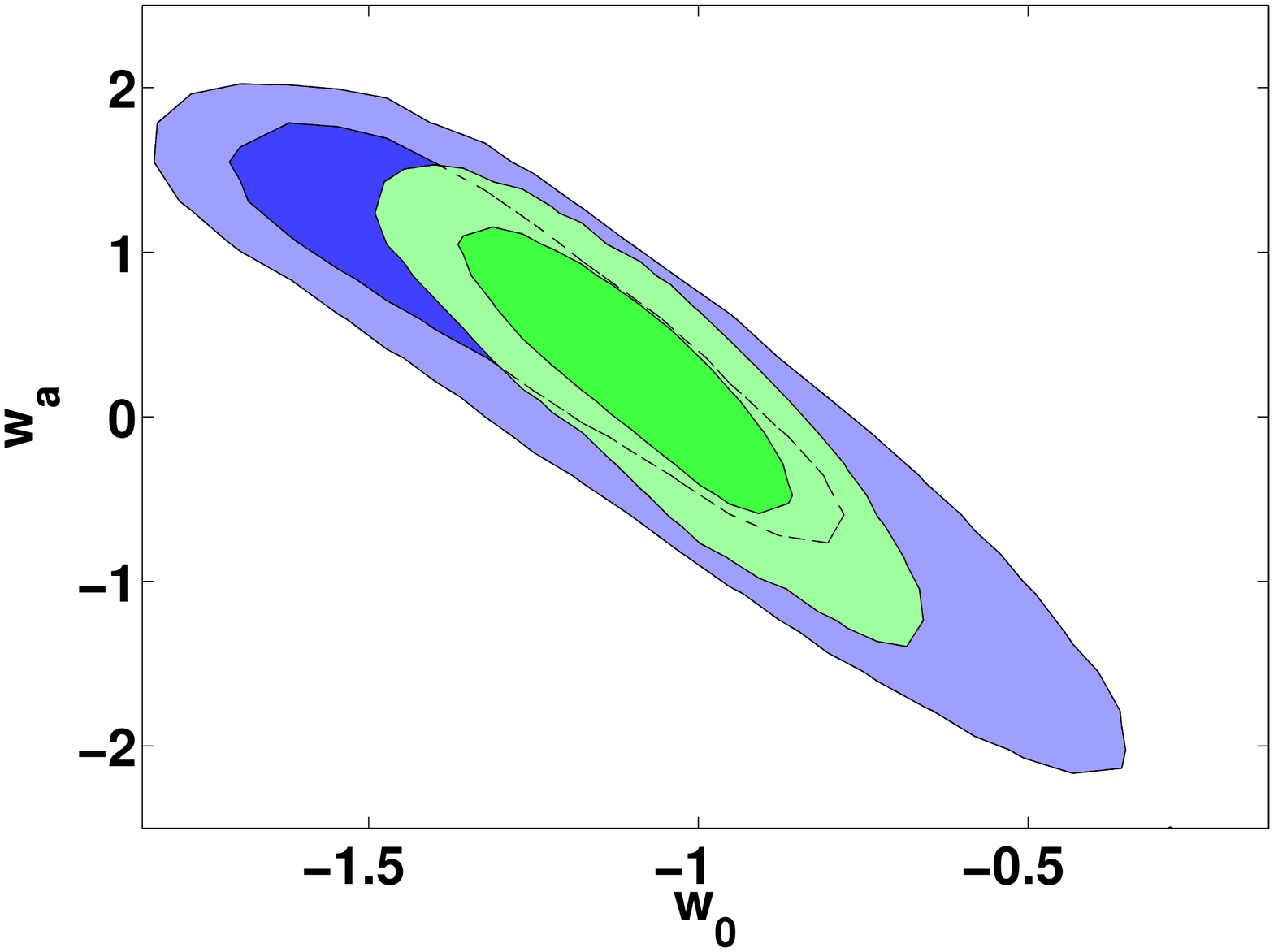}
\caption{Test analysis with $r=0$, evolving DE. Left panel: 1 and 2$\sigma$ contours  $\Omega_m- \Omega_{\Lambda}$ diagram. In blue the combination PolarBear + Planck with 
dynamical DE, in green PolarBear + Planck, with $\Lambda$CDM.   Right panel: 1 and 2$\sigma$ contours for $w_a-w_0$. In blue the results obtained when SNe are not included, in green when SNe data were considered. }
\label{fig:sne}
\end{center}
\end{figure}

\begin{table}[htb]\footnotesize
\begin{center}
\begin{tabular}{c|c|c|c|c|c|c|c}
\hline
\hline
$h^2 \Omega_b$ & $h^2 \Omega_c$ & $100  \cdot \theta$ & $\tau$ & $n_s $ & $A_s$ & $w_0$ &$w_a$\\
\hline
0.02258 & 0.1109 & 1.0388 & 0.087 & 0.963 & 2.43 $\cdot 10^{-9}$ &-1 &0\\
\hline
\hline
\end{tabular}
\caption{{Set of cosmological parameters and adopted values for the cases $r=0$ and $r=0.05$ of simulated data.}}
\label{tab:wmap}
\end{center}
\end{table}
Therefore, two different fiducial models were adopted concerning the amplitude of primordial GWs, corresponding to their absence ($r=0$) and to $r=0.05$. The latter case corresponds 
to a detectable value also in a more realistic case in which data analysis includes foreground cleaning and power spectrum estimation is chained to the MCMCs \citep{stivoli_etal_2010,fantaye_etal_2011}. \\
{Using these sets we compute the fiducial power spectra $C_{\ell}^{i}$ with $i=TT,TE,EE,BB$, in order to compare them with the theoretical models generated by exploring the parameter space. In this work with make use of the {\tt cosmomc} package for that. We add a noise bias to these fiducial spectra, consistently with the mentioned instrumental specifications.} 
For each frequency channel which is listed in Table \ref{tab:exp}, the detector noise considered is $w^{-1} = (\theta\sigma)^2$, where $\theta$ is the FWHM (Full-Width at
Half-Maximum) of the instrumental beam if one assumes a Gaussian and circular profile and $\sigma$ is the sensitivity $\Delta T$. To each of the $C_\ell$ coefficients 
the added contribution from the noise is given by:
$N_\ell = w^{-1}e^{(\ell(\ell+1)/\ell_b^2)}$,
where $\ell_b$ is given by $\ell_b \equiv \sqrt{8\ln2}/\theta$. 
The MCMCs were conducted by adopting a convergence diagnostic based on the Gelman and Rubin statistics \cite{gelmanrubin}. We sample eight cosmological parameters ($\Omega_{b}h^2$, $\Omega_{c}h^2$, 
$\tau$, $n_s$, $A_s$, $\Omega_{\Lambda}$, $z_{reion}$, as well as the Hubble expansion rate 
$H_0$), the $w_0$ and $w_a$ DE parameters, and $r$ adopting flat priors. We make use of priors coming from different probes in the {\tt cosmomc} package, specifically 
Baryon Acoustic Oscillations (BAO) \citep{percival_etal_2011,blake_etal_2011}, Supernovae (SNe) data \citep{kessler_etal_2009}, results from the Hubble Space Telescope (HST) \citep{riess_etal_2009}. \\
In order to calibrate our pipeline, we first consider a $\Lambda$CDM model with $r=0$, varying both the DE parameters $w_0, \, w_a$ or keeping them fixed to a CC through the MCMCs, and considering 
for simplicity the combination of Planck and one sub-orbital experiment (PolarBear). The results in the ($\Omega_{\Lambda},\Omega_{m}$) plane are shown in Fig. \ref{fig:sne} (left panel), 
showing the 1 and 2$\sigma$ contours for the case of a Cosmological Constant (green) and dynamical DE (blue). The decrease in constraining power due to the extra degrees of 
freedom is evident, although the shape of the contour regions is rather stable. Our interpretation is that the introduction of new degrees of freedom affects the precision on the measurement of the two parameters considered. On the other hand, the distance to last scattering is degenerate between cosmological abundances and expansion history, resulting in a geometric degeneracy for the non-lensed pure CMB dataset. Our forecasted datasets contain both CMB lensing measurements, as well as external data on the recent expansion history; we see here how this procedure eliminates such degeneracies. The residual effect is represented by a loss of precision due to the higher dimension of the parameter space, accounting now for a dynamical DE. We further investigate this point in the right panel of Fig. \ref{fig:sne}, where the results in presence (green) or absence (blue) of the SNe measurements are shown, confirming the substantial relevance of external measurements of the expansion history at low redshift, as 
anticipated in earlier works \citep{hu_etal_2006}. 

\begin{table}[htb]
\begin{center}
\begin{tabular}{rccc}
Experiment & Channel & FWHM & $\Delta$T/T\\
\hline
Planck & 70 & 14' & 4.7\\
\phantom{Planck} & 100 & 9.5' & 2.5\\
\phantom{Planck} & 143 & 7.1'& 2.2\\
\phantom{Planck} & 217 & 5.0' & 4.8 \\

$f_{sky}=0.85$ & & & \\
\hline 
EBEX & 150 & 8' & 0.33 \\
\phantom{EBEX} & 250 & 8' & 0.33 \\ 
\phantom{EBEX} & 410 & 8' & 0.33\\ 
$f_{sky}=0.01$ & & & \\

\hline 
PolarBear & 90 & 6.7' & 0.41 \\
\phantom{PolarBear} & 150 & 4.0' & 0.62 \\ 
\phantom{PolarBear} & 220 & 2.7' & 2.93\\ 
$f_{sky}=0.03$ & & & \\

\hline
CMBpol & 70 & 12' & 0.148\\
\phantom{CMBpol} & 100 & 8.4' & 0.151\\
\phantom{CMBpol} & 150 & 5.6'& 0.177\\

$f_{sky}=0.85$ & & & \\

\hline
\hline
\end{tabular}
\caption{Planck, EBEX, PolarBear and CMBpol performance specifications. Channel frequency is given in GHz, beam FWHM in arcminutes, and the sensitivity for $T$ per pixel in $\mu$K/K. 
The polarization sensitivity for both $E$ and $B-$modes is $\sqrt{2}\Delta T/T$.}
\label{tab:exp}
\end{center}
\end{table}

{It is interesting to compare the present case in which lensing $B-$modes are probed directly by CMB sub-orbital experiments with the case in which the lensing is extracted from all sky CMB anisotropy maps as expected by adopting the nominal performance of operating (Planck) and proposed post-Planck polarisation dedicated CMB satellites (see CMBpol and the COsmic Origin Explorer (CORE) \cite{CMBpol,core}); the latter cases will give us an estimate of the improvement in the constraining power on $w_0, w_a$ as a function of the satellite instrumental specifications. A similar approach has already been applied to the $TT$ spectrum by the SPT collaboration in \cite{sptlensing}; the case for this analysis is different since the focus is set on the $B-$modes. We create {simulated datasets for Planck and CMBpol}, adopting nominal performances as in the previous case, but adding the forecasted lensing potential measurements. Our aim is to quantify, {in these cases}, the efficiency on determination of the expansion parameters $w_{0}$ and $w_{a}$, {and how they scale with satellite instrumental capabilities, reaching cosmic variance limit also for polarization as in the cases of planned post-Planck satellite CMB experiments}; therefore we keep $r=0$ fixed and let the CPL parameters vary. We use the lensing extraction method presented in \cite{lensextr} where the authors construct the weighting factor $W$ of Eq. (\ref{eq:estimator}) as a function of CMB power spectra $C_{ab}$, with $ab = TT,\, TE,\, EE,\, EB,\, TB$. The $BB$ spectrum is excluded because the adopted method is only valid when the lensing contribution is negligible compared to the primary anisotropies; this assumption fails for $B-$modes, which are not considered in this analysis, by modifying {\tt cosmomc} according with \cite{Perotto:2006rj}. This aspect, as well as the instrumental sensitivity, implies that lensing measurements in this case come mainly from sub-degree $T$ and $E$ anisotropy data. We study the constraining power on CPL parameters from Planck data in three cases: first, when lensing measurements are used, second, without lensing, but with the inclusion of the priors introduced above (BAO, HST, SNe), and finally using both. {We performed this analysis also on a CMBpol-like experiment using the specifications in \cite{CMBpol}; the major uncertainty on the data from such an experiment will be due to cosmic variance.} Results are presented in Table \ref{tab:lensextr}.\\
\begin{table}[h]
\begin{center}
\begin{tabular}{r|c|c|c}
Planck & CMB+lensing extraction & CMB+priors & CMB+lensing extraction+priors \\
\hline
\hline
 $\Delta(w_0)$ & $0.5$ & $0.2$ & $0.2$ \\
$\Delta(w_a)$ & $1.1$ & $0.6$ & $0.6$ \\
\hline
\hline
CMBpol & CMB+lensing extraction & CMB+priors & CMB+lensing extraction+priors \\
\hline
\hline
 $\Delta(w_0)$ & $0.4$ & $0.159$ & $0.150$ \\
$\Delta(w_a)$ & $1.0$ & $0.57$ & $0.497$ \\
\hline
\hline
\end{tabular}
\caption{$1\sigma$ uncertainties on CPL parameters $w_0,\, w_a$ for Planck {and for a CMBpol specifications}  when using lensing extraction, when using external priors and when combining both, in the case $r=0$.}
\label{tab:lensextr}
\end{center}
\end{table}
{Let us focus first on the comparison between CMB satellite lensing measurements and the case in which the lensing is probed through the lensing dominated part of the $B-$mode spectrum.} 
As it can be seen comparing with the contours in Figure \ref{fig:sne}, the relevance of lensing measurements is comparable {in the two cases}; moreover, it is found that the priors have a comparable relevance. We conclude that satellite lensing measurements using $T$ and $E$, and sub-orbital ones directly accessing lensing $B-$modes, have a comparable capability for constraining the expansion history. Both cases are relevant to study, as the impact of non-idealizations including systematics as well as removal of foreground emissions may produce different outcomes \cite{perotto2009,fantaye2012}. {Let us now discuss the differences between the case of Planck, which is a cosmic variance limited experiment for total intensity, with respect to the enhanced capability of planned post-Planck satellites, approaching the same limit for polarization as well. As the results show, the improvement in the instrumental specification does cause an enhancement of the constraining capability corresponding to a factor $20\%$ for $w_0$ and $10\%$ for $w_a$; when priors are considered, the results improve by a factor of about $6\%$ for $w_0$ and $15\%$ for $w_a$. We conclude that the improvement is sensible but does not change the order of magnitude of the forecasted precision, and we argue that this is consistent with the fact that Planck is cosmic variance limited in total intenisity, which is the dominant part of the CMB anisotropy signal.}\\ 
In the following we focus on the capability of constraining the expansion parameters using the $B$-modes, in order to study if new degeneracies arise when the relative amplitude between PGWs (through variations of $r$) and the lensing spectrum (as traced by lensing $B-$modes) vary at the same time.\\}

\section{Results}
\label{sec:results}

We study here the recovery of the primordial tensor to scalar ratio, performed while  varying the cosmological expansion history. As we already pointed out, we consider two cases, for a null 
($r=0$) and positive ($r=0.05$) detection. In both cases, the fiducial DE model is $\Lambda$CDM, and the generalized expansion history is parametrized by $w_{0}$ and $w_{a}$. In order to verify 
the relevance of sub-orbital probes, probing the lensing peak in the $B-$mode spectrum, we consider the case of pure satellite CMB data separately from the one with joint satellite and sub-orbital 
probes. 

The results on $r$ as 2$\sigma$ upper limits and 1$\sigma$ statistical uncertainties in the null and positive detection cases respectively, as well as the corresponding constrains on CPL parameters are shown in Table \ref{tab:res}. {In the case with a non-vanishing fiducial value of $r$, a change in the MCMC recovered value of $r$ is present when the theoretical model or the experimental configuration are changed. In order to address the reason of the differences in the recovered mean value
of $r$ we computed the Gelman and Rubin indicator for the chains we performed,
finding that the differences we see can be ascribed to fluctuations in the MCMC
procedure (see e.g. \cite{method} for a more specific discussion on this topic). Nevertheless, note that}, as expected, the results obtained by adopting the nominal specifications of Planck are in agreement with \cite{Efstathiou:2009xv} for $\Lambda$CDM. A first result concerns the quantification of precision loss of the recovery on $r$ when a generalized expansion rate is considered, and when only satellite CMB data are considered. 
This corresponds roughly to $10\%$ for the null and about $5\%$ for positive detections of $r$. The interpretation is related to the extra degrees of freedom considered, while as in the previous 
sections, the lensing component of simulated spectra, as well as the priors on the expansion history from external probes, help reducing geometric degeneracies, leaving room only 
for an increase in the statistical error of the various measurements, which we quantify here. It is interesting now to look at the case when all the CMB probes are considered, verifying that 
the precision loss in this case falls below a detectable level. This result is uniquely related to the enhanced sensitivity of sub-orbital probes, allowing for a deeper study of  the lensed component 
of CMB spectra, and in particular on the lensing peak in $B-$modes. Concerning the CPL parameters ($w_0,\, w_a$), it is possible to see in Table \ref{tab:res} how the constraints 
do not degrade switching from the $r=0$ to the $r=0.05$ simulated dataset. This shows, as previously stated, that there are no detectable degeneracies between $r$ and CPL parameters 
in our considered datasets. 
Moreover we can also notice how constraints on ($w_0$, $w_a$) do not improve much if we use sub-orbital
experiments alongside satellite data to get better CMB sensitivity; this highlights the fact that the prior 
we used, most of all the SNe data, are crucial to constrain DE quantities. 
\begin{table}[h]
\begin{center}
\begin{tabular}{r|c|c}
Experiments, fiducial & $r=0$ & $r=0.05$\\
\hline
\hline
Planck with priors, $\Lambda$CDM & $r< 0.029$ & $r = 0.057 \pm 0.022$\\
Planck with priors, CPL & $r<0.031 $ & $r= 0.059 \pm 0.023$\\
all experiments, $\Lambda$CDM  & $r < 0.025$  & $r= 0.057 \pm 0.020$ \\
all experiments, CPL  & $r<0.025$ & $r = 0.056\pm 0.020$ \\
\hline
Planck with priors, CPL & $w_0=-1.1\pm0.2 $ & $w_0=-1.1 \pm 0.2$\\
all experiments, CPL  & $w_0=-1.1\pm 0.2$ & $w_0=-1.1\pm0.2$ \\
\hline
Planck with priors, CPL & $w_a=0.3\pm 0.6 $ & $w_a=0.3\pm0.6$\\
all experiments, CPL  & $w_a=0.3\pm0.6$ & $w_a=0.2\pm 0.6$ \\
\end{tabular}
\caption{$2\sigma$ upper limits and $1\sigma$ uncertainties for the measurements of $r$ for the null and positive detection cases, and $1\sigma$ uncertainties for the measurements of the CPL parameters $w_0,\, w_a$ for the different expansion models and dataset combinations.}
\label{tab:res}
\end{center}
\end{table}

\begin{figure}[ht]
\centering
\includegraphics[width=6cm]{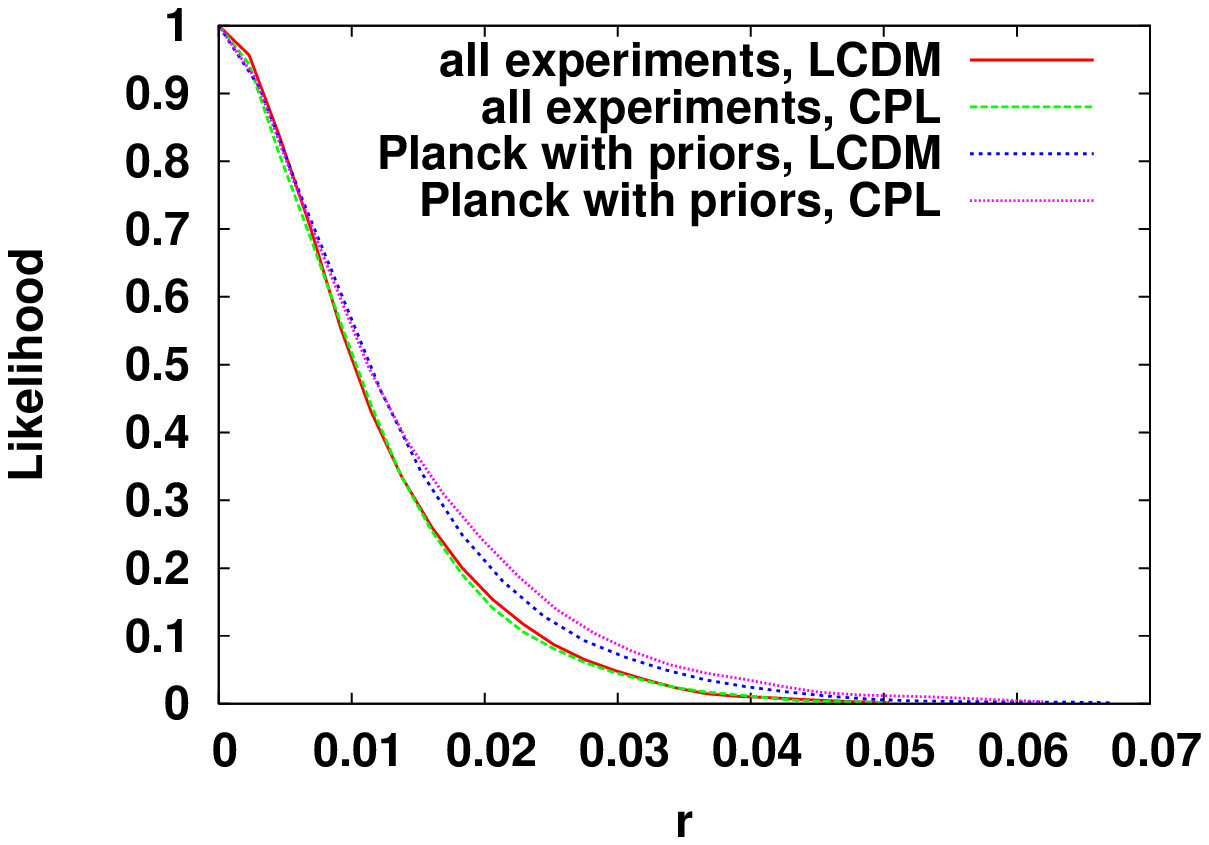}
\includegraphics[width=6cm]{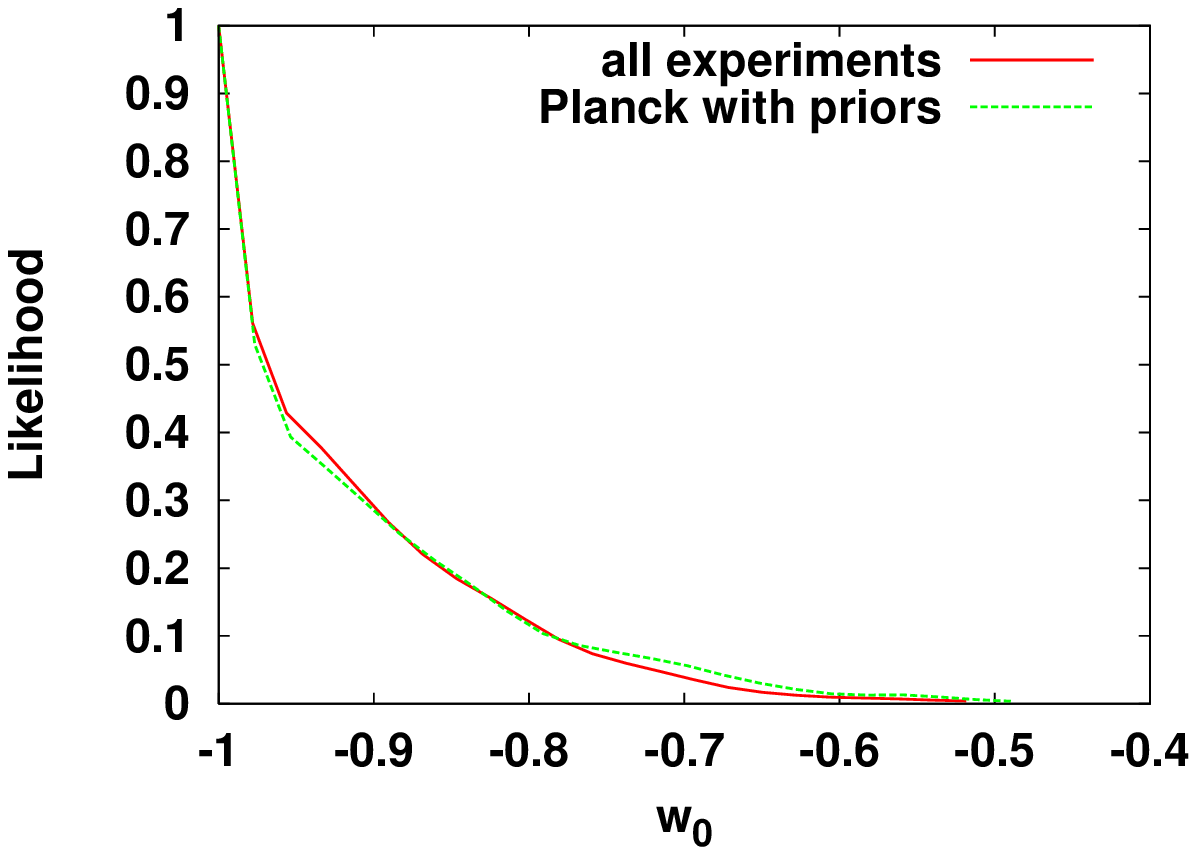}
\includegraphics[width=6cm]{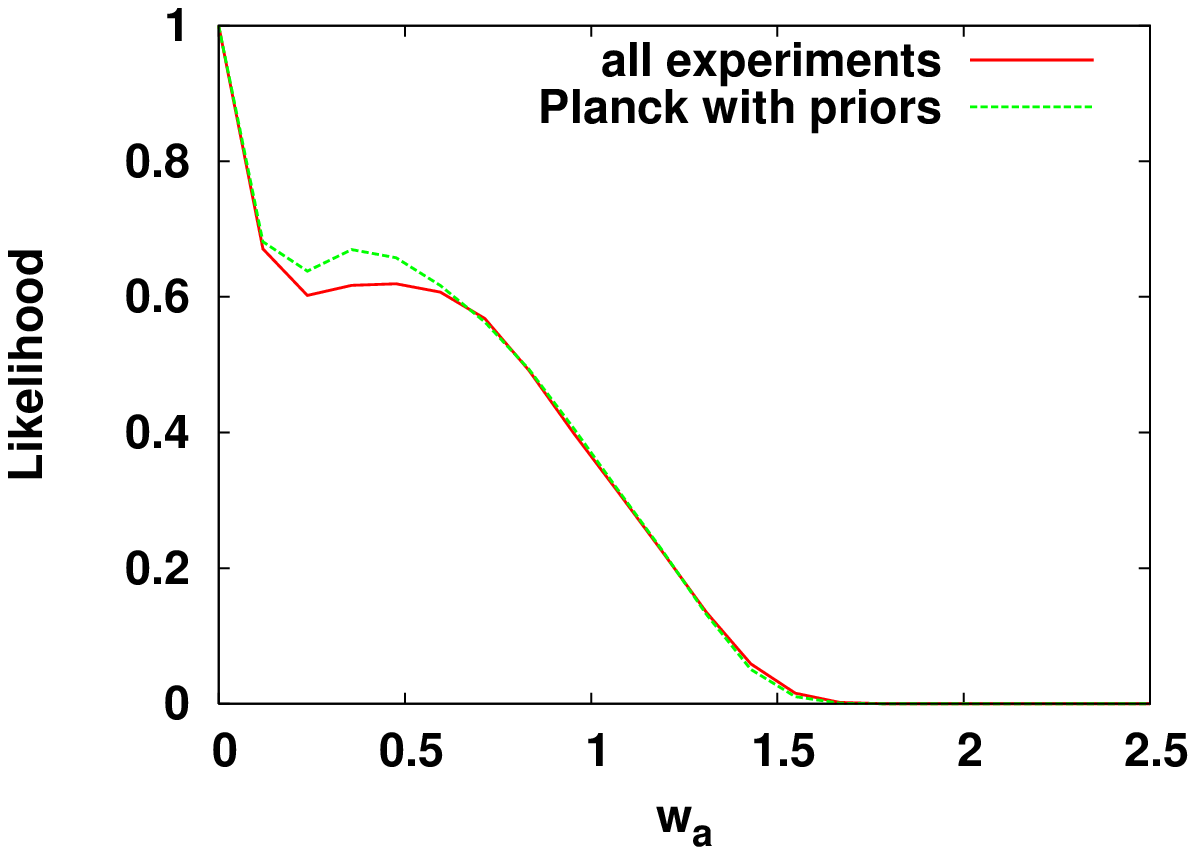}
\caption{One-dimensional contours for $r$, $w_0$ and $w_a$ respectively, in the case of null detection for $r$; all plots show differences when using satellite, or all CMB datasets; 
the plot for $r$ also includes the $\Lambda$CDM cases.}
\label{fig:monodim}
\end{figure}

\begin{figure}[ht]
\centering
\includegraphics[width=7.5cm,height=5cm]{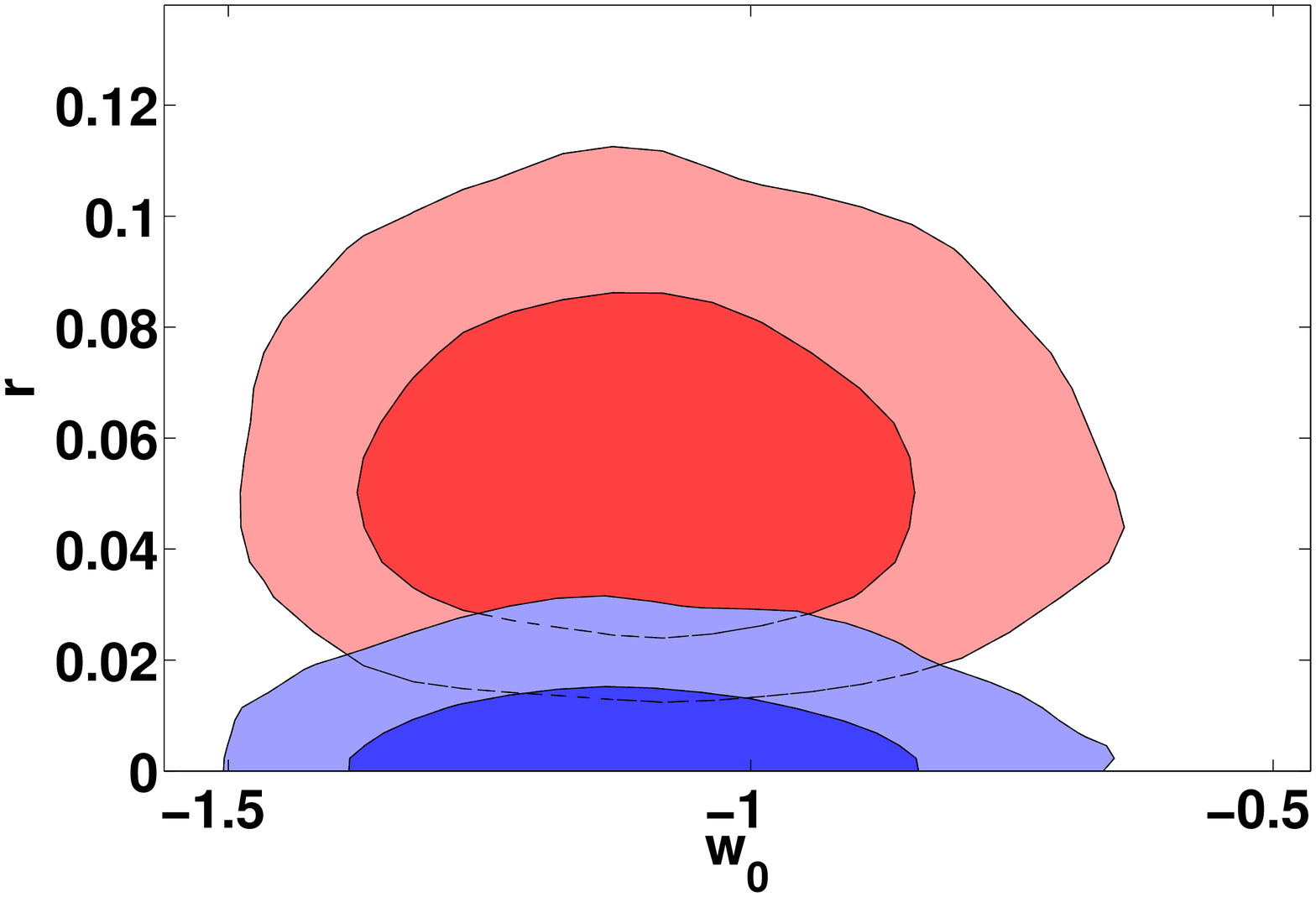}
\includegraphics[width=7.5cm,height=5cm]{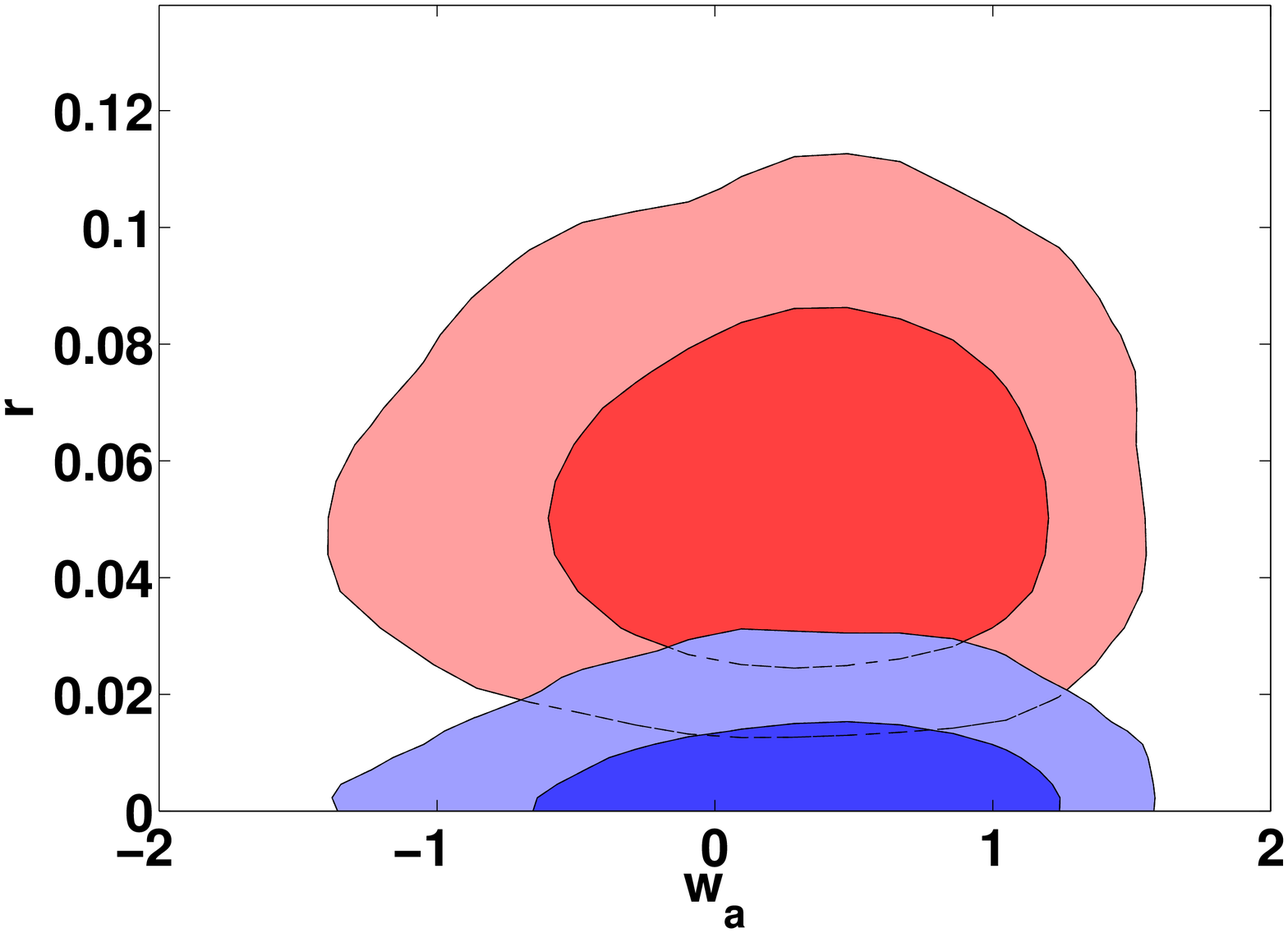}
\includegraphics[width=7.5cm,height=5cm]{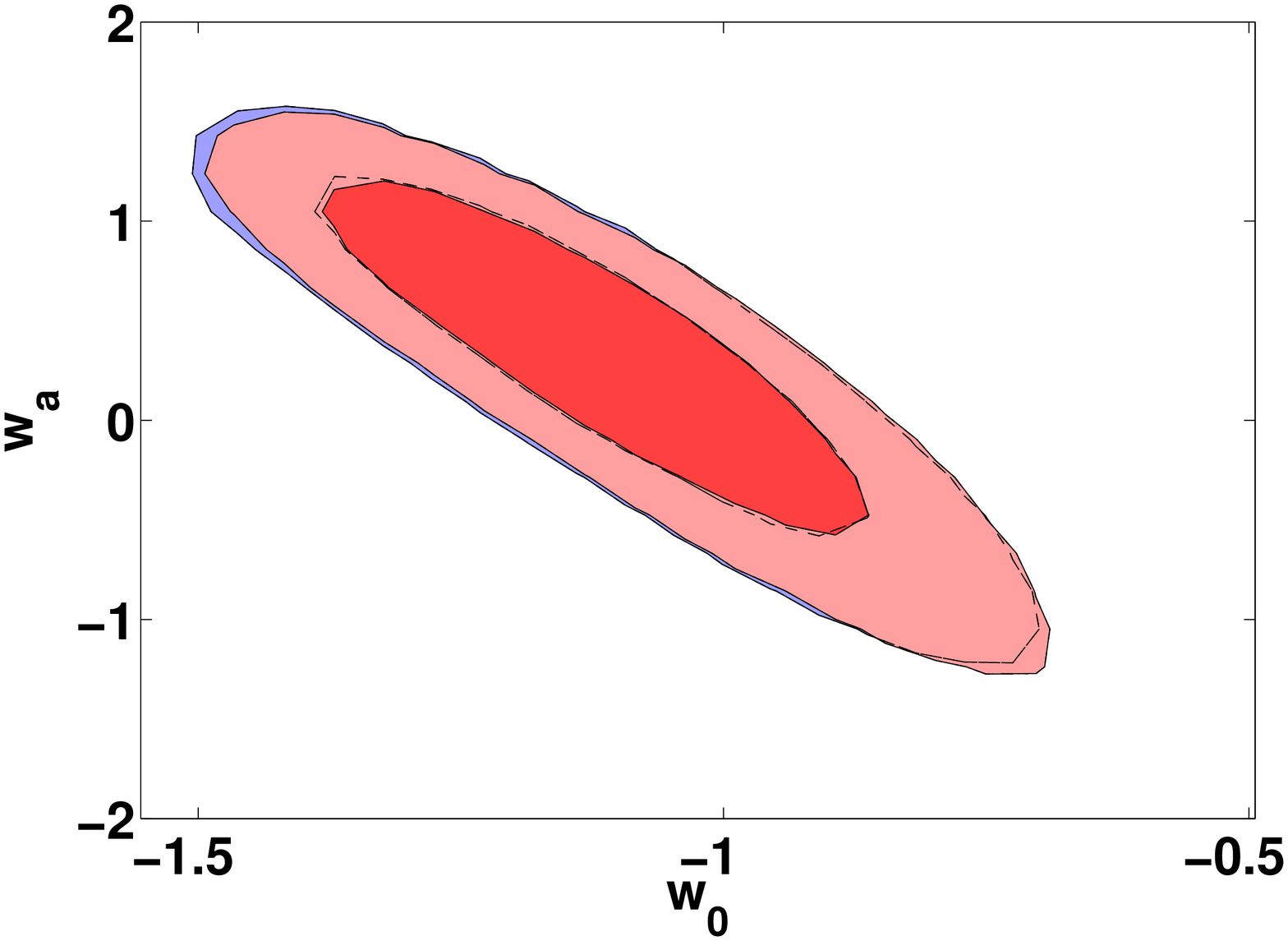}
\caption{Up, left panel: 1 and 2$\sigma$ contours $w_0-r$ diagram for the combination of all considered datasets. In blue, the upper limit on the simulated data with fiducial value $r=0$. In red, the 
case of simulated data with fiducial value $r=0.05$. Up, right panel: 1 and 2$\sigma$ contours $w_a-r$ diagram. In blue, the upper limit on the simulated data with fiducial value $r=0$. In red, the 
case of simulated data with  fiducial value $r=0.05$.
Lower panel: 1 and 2$\sigma$ contours $w_0-w_a$ diagram. In blue, the constraints on the simulated data with fiducial value $r=0$. In red, the 
case of simulated data with fiducial value $r=0.05$. }
\label{fig:detectionr}
\end{figure}

These limits have been derived from one-dimensional contours, which are shown in Fig. \ref{fig:monodim}, reporting the null detection case only, for simplicity, for $r$ and the DE 
parameters, and restricting to the case of DE models with $w>-1$; it can be noticed how considering the whole CMB datasets yields an improvement on the detection of $r$, reflecting 
Table \ref{tab:res}, while almost no difference is noticeable between the cases of dynamical DE or $\Lambda$. 
{ Looking at the first panel in Fig. \ref{fig:monodim} one can in particular appreciate how the shape in the likelihood for $r$ is non-Gaussian, justifying our choice of going through a MCMC analysis rather than relying on a Fisher matrix approach.} For DE parameters, we notice no particular improvement in considering the case of all CMB or pure satellite datasets alongside SNe, BAO and HST data. The same holds when looking at two-dimensional contours, 
shown in Fig. \ref{fig:detectionr} in the ($r$,$w_{0}$), ($r$,$w_{a}$) and ($w_0$,$w_{a}$) planes, for the null (blue) and positive (red) detection cases: in none of the three panels 
a significant improvement in DE parameter recovery is shown, even allowing for cosmologies with $w<-1$. We also notice that no degeneracies among these parameters are detectable with the datasets we consider. The figures also quantify the precision achievable on DE parameters, being comparable and of the order of a few ten percents, for both parameters and both fiducial models. 

\begin{figure}[ht]
\begin{center}
\begin{tabular}{cc}
\includegraphics[width=6cm,height=4cm]{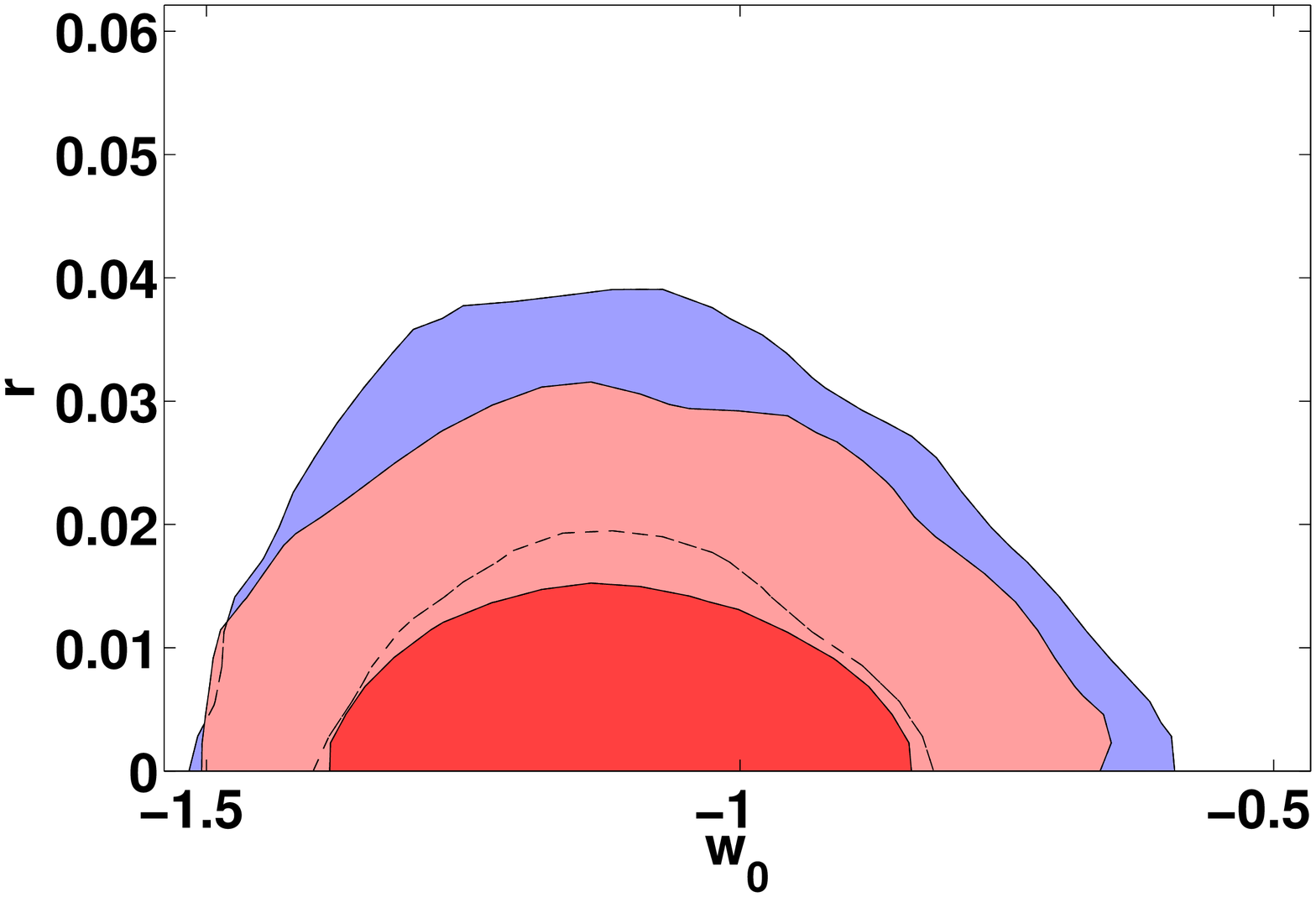}&
\includegraphics[width=6cm,height=4cm]{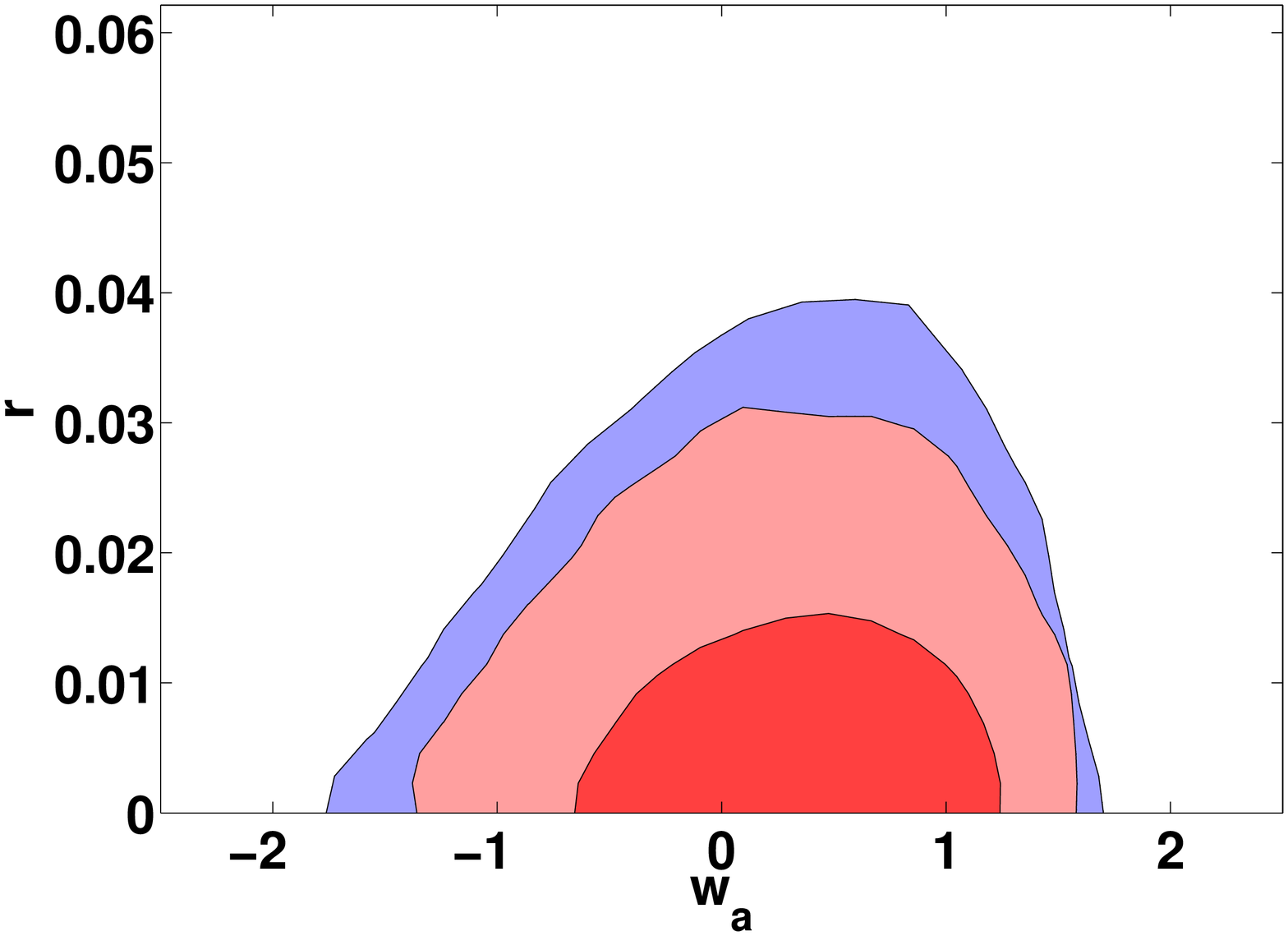}\\
\includegraphics[width=6cm,height=4cm]{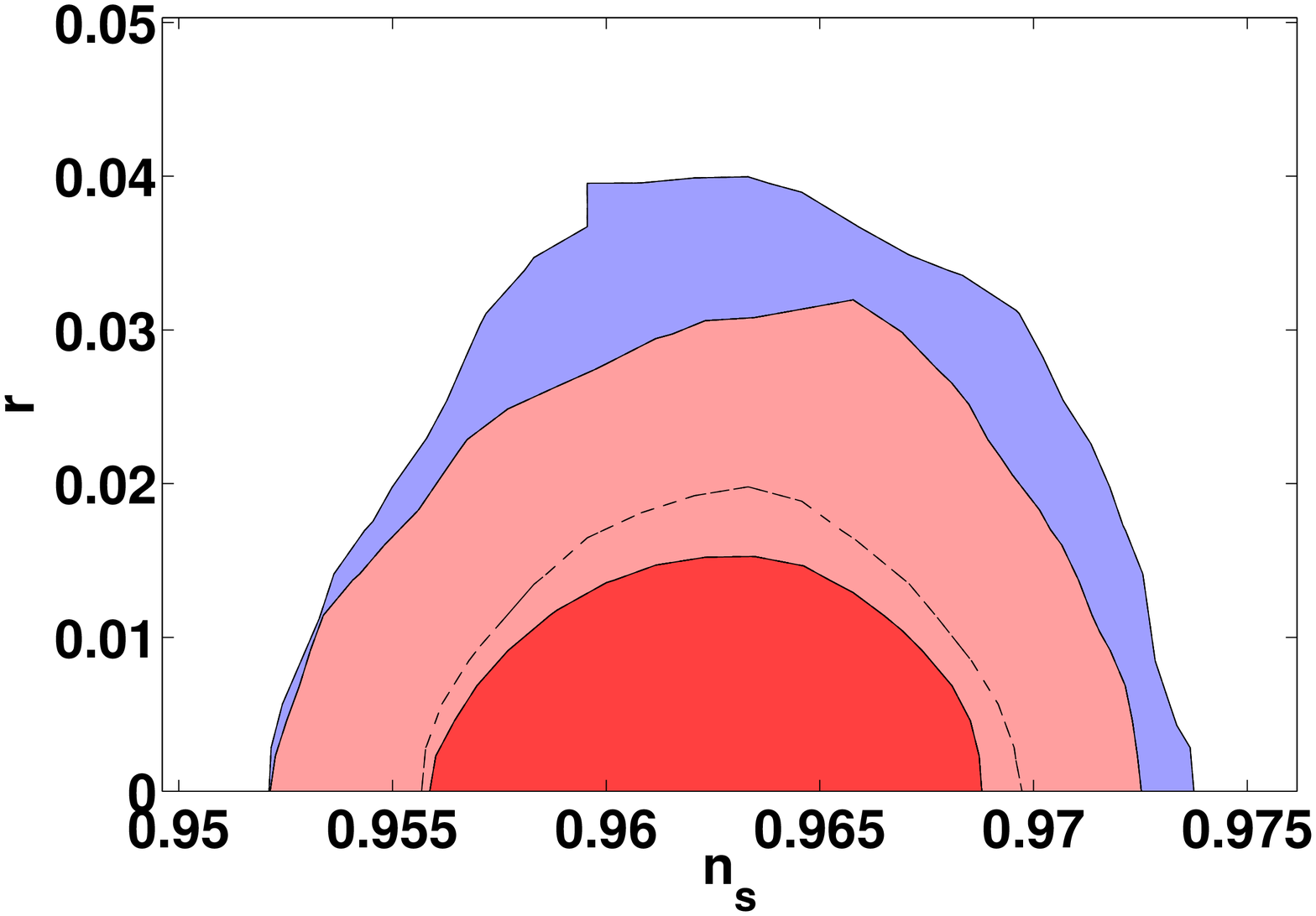}&
\includegraphics[width=6cm,height=4cm]{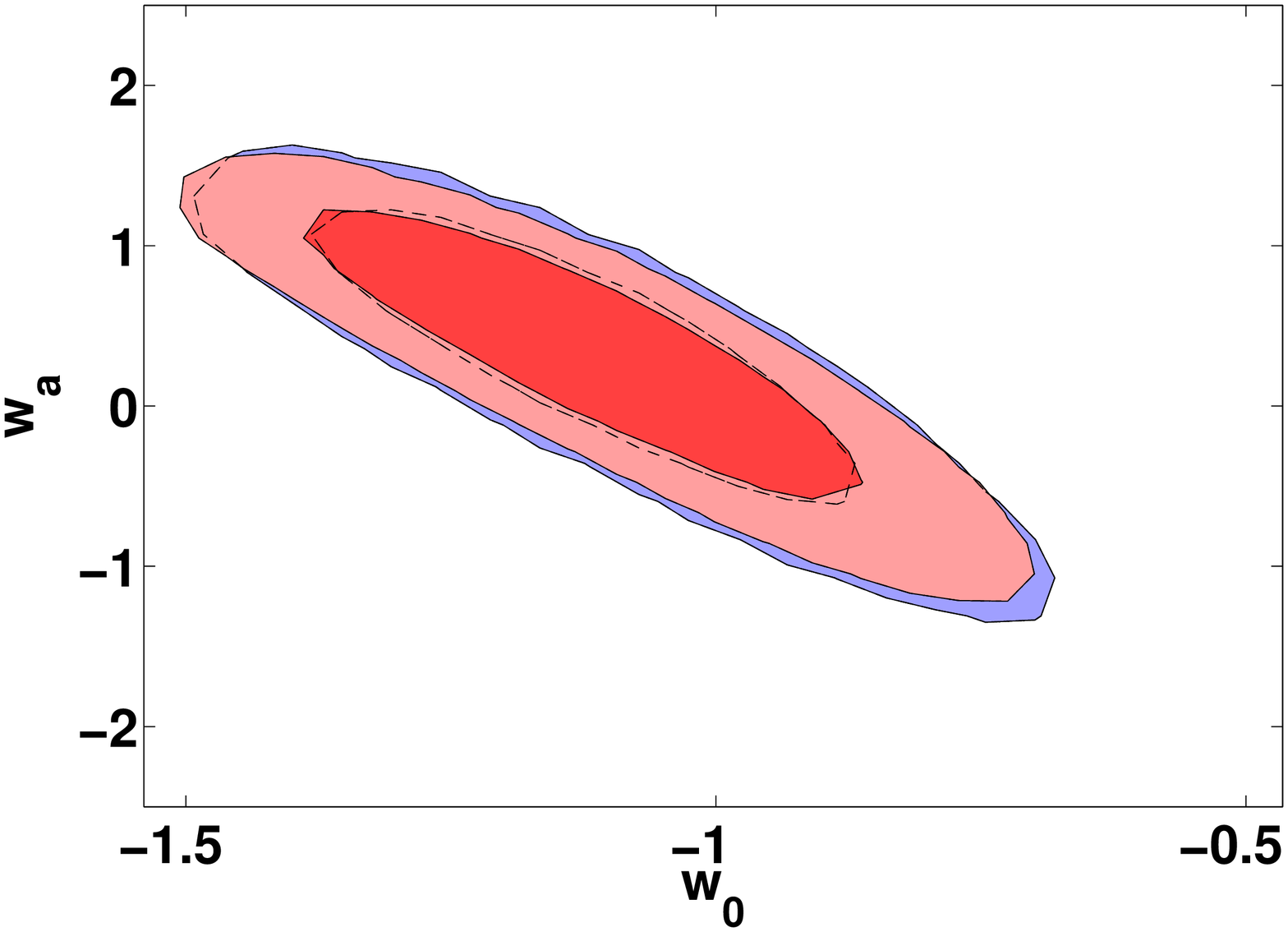}
\end{tabular}
\caption{Results from the analysis on the $r=0$ fiducial value simulated dataset. In all plots, blue contours represent pure satellite CMB data, while the red ones include sub-orbital ones as well. 
From left to right, from top to bottom.
1.  $1-2\sigma$ contours for $r-w_0$. 2. $1-2\sigma$ contours for $r-w_a$. 3. $1-2\sigma$ contours for $r-n_s$ for dynamical DE.  4. $1-2\sigma$ contours for $w_0-w_a$.}
\label{fig:allresults1}
\end{center}
\end{figure}

\begin{figure}[ht]
\centering
\includegraphics[width=6cm,height=4cm]{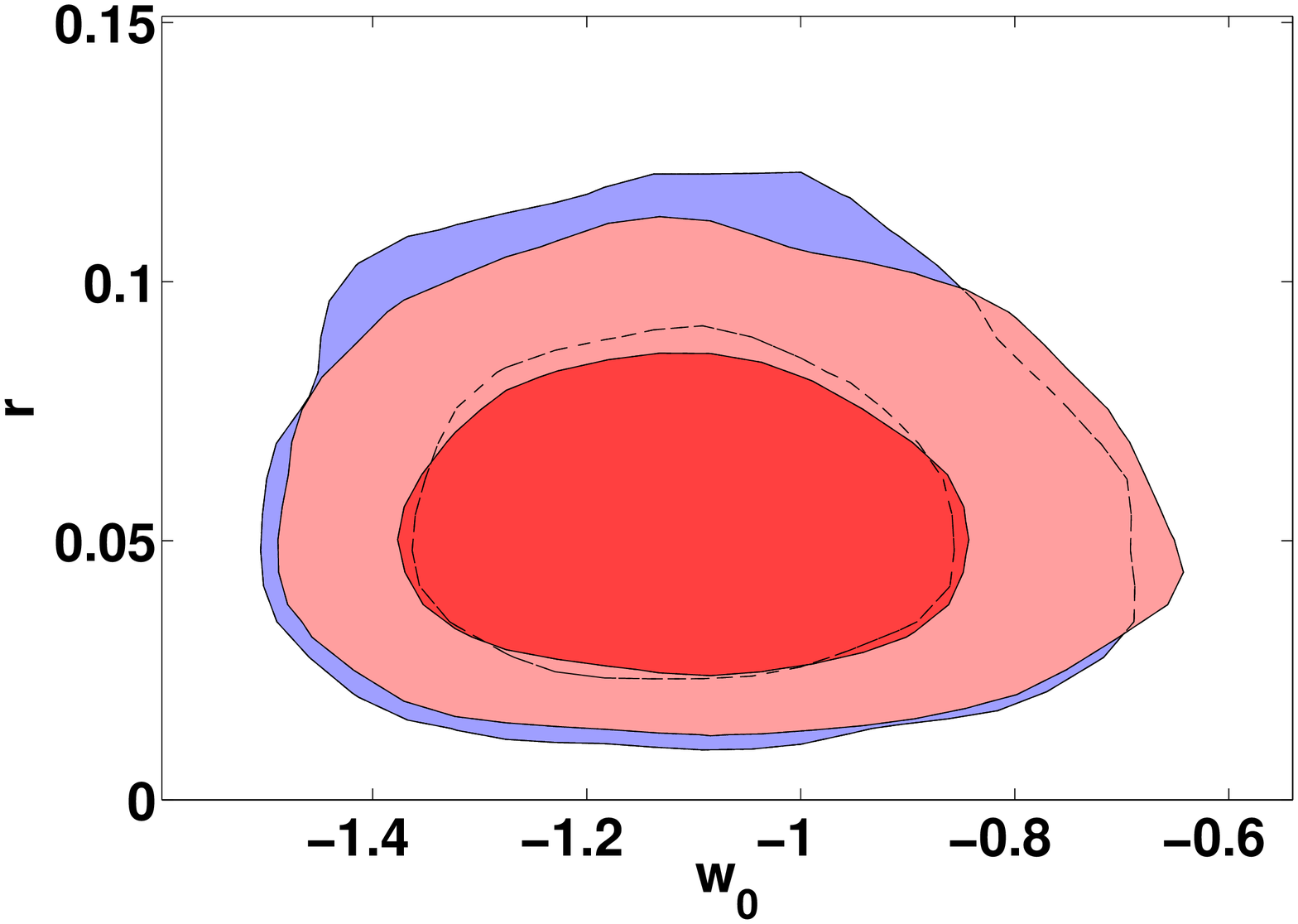}
\includegraphics[width=6cm,height=4cm]{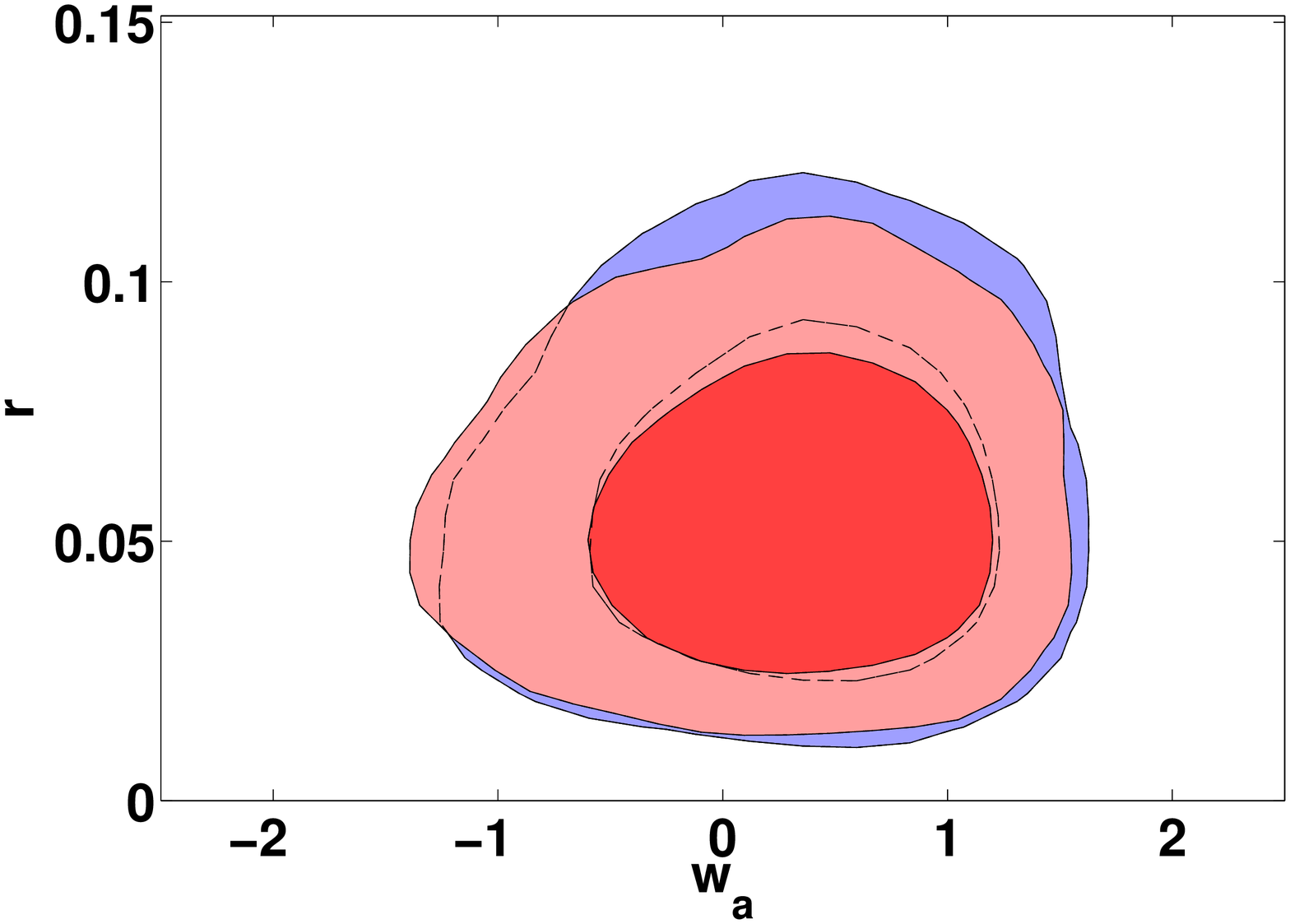}
\includegraphics[width=6cm,height=4cm]{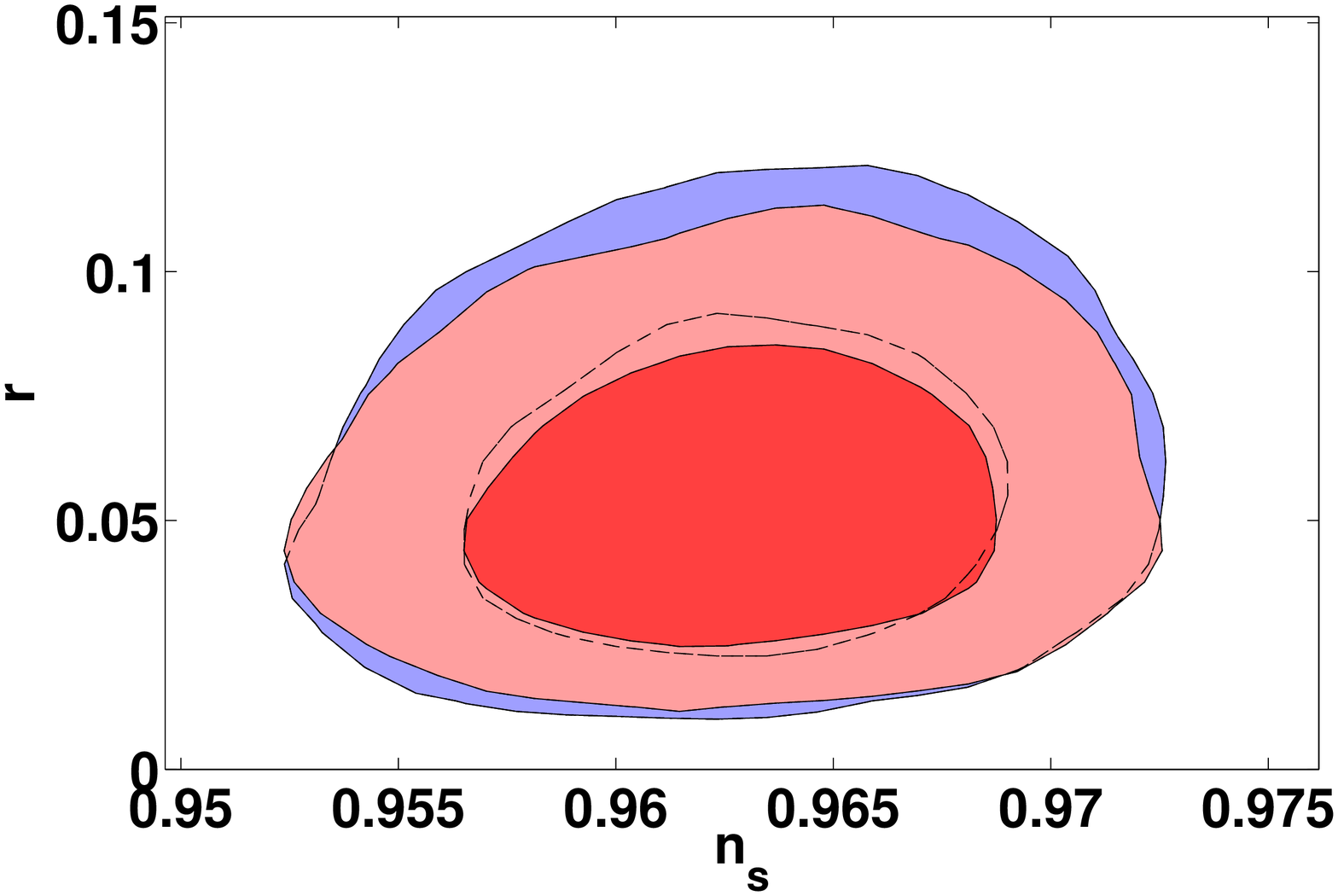}
\includegraphics[width=6cm,height=4cm]{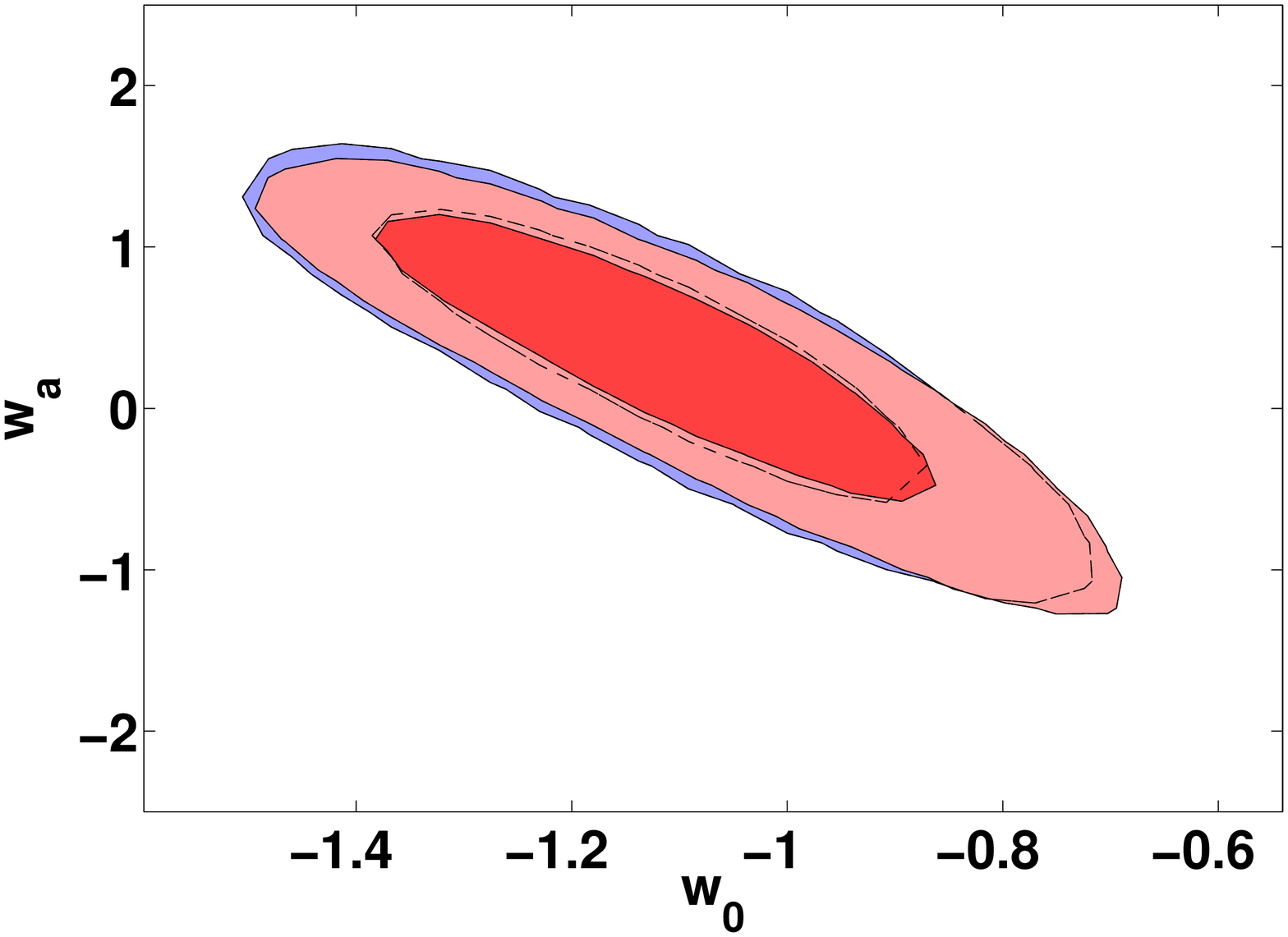}
\caption{Results from the analysis on the $r=0.05$ fiducial value simulated dataset. In all plots, blue contours represent pure satellite CMB data, while the red ones include sub-orbital ones as well. 
From left to right, from top to bottom.
1.  $1-2\sigma$ contours for $r-w0$. 2. $1-2\sigma$ contours for $r-w_a$. 3. $1-2\sigma$ contours for $r-n_s$ for dynamical DE.  4. $1-2\sigma$ contours for $w_0-w_a$.}
\label{fig:allresults2}
\end{figure}

Finally, we show other relevant two-dimensional contour plots for the case of null detection (Fig. \ref{fig:allresults1}) and for the $r=0.05$ fiducial value (Fig. \ref{fig:allresults2}), 
highlighting how with the data considered here it is not possible to detect any degeneracy between the primordial tensorial mode parameter $r$  and other cosmological parameters. Despite this remarkable result, we stress that our results concern a nominal performance of the various datasets, and in particular do not consider foreground cleaning or other systematic effects, 
which were pointed out as possible sources of bias for $r$ in previous works \citep{stivoli_etal_2010,fantaye_etal_2011,Pagano:2009kj}.

\section{Concluding remarks}
\label{sec:discussion}

The primordial Gravitational Waves (PGWs) and lensing power constitute the dominant effects for the $B-$mode polarization in 
the anisotropies of the Cosmic Microwave Background (CMB). While the former is dominated by the physics of the early Universe, 
parametrized through the primordial tensor-to-scalar ratio $r$, the latter is instead due to structure formation, and thus influenced by
the expansion rate at the epoch of the onset of cosmic acceleration. This, in turn, is dependent on the underlying  dynamics of 
the Dark Energy (DE). Despite both signals being present in the CMB $B-$modes, their joint measurement in terms of parameter estimation
was never considered, and this work represents a first step in this direction. 

{ We first address the lensing relevance for constraining our parametrization of the expansion history, assuming no PGWs. We find comparable results when the lensing is extracted from $T$ and $E$ data and when the lensing is traced directly through lensing $B-$modes, by forthcoming satellite and sub-orbital data, respectively, {both for a Planck experiment and for a CMBpol experiment}. Focusing on the latter case, where the two processes directly compete for detection in $B-$modes, we quantify} the constraining power on the abundance of 
PGWs which is expected from combined forthcoming satellite and sub-orbital experiments probing CMB polarization in cosmologies with 
generalized expansion histories, parametrized through the present and first redshift derivative of the DE equation of state, $w_{0}$ and $w_{a}$, 
respectively. We find that in the case of pure satellite measurements, corresponding to the Planck nominal performance, the constraining power on 
GWs power is weakened by the inclusion of the extra degrees of freedom, resulting in an increase of about $10\%$ of the upper limits on $r$ in fiducial 
models with no GWs, as well as a comparable increase in the error bars in models with non-zero tensor power.
The inclusion of sub-orbital CMB experiments, 
capable of mapping the $B-$mode power up to the angular scales which are affected by lensing, has the effect of making such loss of constraining power 
vanishing below a detectable level. We interpret these results as a joint effect of the CMB and external datasets:  the former are able, in particular with 
the data from sub-orbital probes, to access the region of $B-$modes which is lensing dominated, and therefore sensitive to the DE abundance at the onset of acceleration; 
the latter, as the case of Type Ia SNe and the Hubble Space Telescope, are on the other hand strongly constraining the dynamics of cosmic expansion at present.
By inspecting the constraints on all cosmological parameters, including those parametrizing the expansion history, we also show that the datasets we consider do not highlight new degeneracies in the parametrization we consider. 

Our results indicate that the combination of satellite and sub-orbital CMB data, with the available external data useful to inquire the late time expansion 
history, can be used for constraining jointly the dynamics of the DE as well as the primordial tensor-to-scalar ratio, with no new degeneracies or 
significant loss of sensitivity in particular on $r$ with respect to the case in which a pure Cosmological Constant determines the late time cosmological 
expansion. Our assumptions of course include the nominal performance of these experiments, and no realistic data analysis consisting in the inclusion of 
foregrounds in the CMB data, as well as systematic errors. It would be interesting to further investigate this phenomenology in specific DE models, and considering 
the role of future surveys in giving more accurate constraints.
\newpage
\section*{Acknowledgements}

This work was supported by the INFN PD51 initiative. MM wants to thank Erminia Calabrese for useful computational discussions and information. CA thanks Fabio Noviello for his comments and useful discussion.
CB also acknowledges support by the Italian Space Agency through the ASI contracts Euclid-IC (I/031/10/0).

\end{document}